\documentclass[twocolumn,trackchanges]{aastex61}

\shorttitle{EXPANSION OF HOT PLASMA INTO COLD PLASMA} \shortauthors{Ben\'a\v{c}ek and Karlick\'y}
\begin{document}

\title{EXPANSION OF HOT PLASMA WITH KAPPA DISTRIBUTION INTO COLD PLASMA}

\author[0000-0002-4319-8083]{Jan Ben\'a\v{c}ek}
\affil{Department of Theoretical Physics and Astrophysics, Masaryk University,
   Kotl\'a\v{r}sk\'a 2, CZ-61137 Brno, Czech Republic}
\email{jbenacek@physics.muni.cz}

\author[0000-0002-3963-8701]{Marian Karlick\'y}
\affil{Astronomical Institute of the Academy of Sciences of the Czech Republic,
CZ-25165 Ond\v{r}ejov, Czech Republic}

\begin{abstract}
The X-ray emission of coronal flare sources can be explained considering the
kappa electron distribution. Motivated by this fact, we study the problem of
how hot plasma with the kappa distribution of electrons is confined in these
sources. For comparison, we analyze the same problem but with the Maxwellian
distribution. We use a 3-D particle-in-cell code, which is large in
one direction and thus effectively only one-dimensional, but describing all
electromagnetic effects. In the case with the Maxwellian distribution, and in
agreement with the previous studies, we show a formation of the double layer at
the hot-cold transition region that suppresses the flux of hot electrons from
hot plasma into the cold one. In the case with the kappa distribution, contrary
to the Maxwellian case, we found that there are several fronts with the double
layers in the hot-cold transition region. It is caused by a more extended tail
in the kappa case than in the Maxwellian one. The electrons from the extended
tail freely escape from the hot plasma into a cold one. They form a beam which
generates the return current and also Langmuir turbulence, where at
some locations Langmuir waves are accumulated. At these locations owing to the
ponderomotive force, Langmuir waves generate density depressions, where the
double layers with the thermal fronts, suppressing the hot electron flux, are
formed. We also show how protons are accelerated in these processes. Finally,
we compared the kappa and Maxwellian cases and discussed how these processes
could be observed.
\end{abstract}


\section{Introduction}

In solar flares, there are very hot plasma sources at some locations, e.g., the
loop-top~\citep{Kolomanski2007} or above-loop-top \citep{Masuda1994,
Krucker2007,Krucker2010} sources that exist for a longer time than the transit
time of hot electrons in these sources. We note that the above-loop-top sources
can be described as the plasmoids located in the rising magnetic rope
\citep{Karlicky2020} or secondary ropes formed in the current sheet below the
rising rope by the plasmoid instability \citep{2007PhPl...14j0703L,Barta2011}.
In both the loop-top and above-loop-top sources, the hot plasma is naturally
confined in the direction perpendicular to the magnetic field lines of the
magnetic loop or ropes (loops with the helical magnetic field and electric
current). However, the question arises how the hot plasma is trapped in the
direction parallel to the magnetic field lines. In the papers by
\cite{Brown79,Arber09,Karlicky15}, it was proposed that it could be caused by
the so-called thermal conduction front. Such thermal fronts have also been
proposed in the interpretation of some observed features in solar flares
~\citep{Farnik83,Rust85,Mandrini96}.

The problem of hot plasma confinement was also studied in the papers by
~\cite{Li2012,Li2013,Li2014,Roberg2018,2019PhPl...26g2103G,Sun19}, where the authors, using
particle-in-cell (PIC) simulations, presented details of the heat flux
suppression at the contact region between hot and cold plasmas.
\cite{Li2012,Li2013,Li2014} showed that at the beginning of the hot plasma
expansion into a cold one, the hot electrons, escaping from the hot plasma region,
trigger the return current, which is unstable due to the electron-ion streaming
(Buneman) instability. During this process, the double layer with the electric
potential jump is formed. The double layer grows over time and supports a
significant drop in temperature and hence reduces heat flux between the hot and
cold regions. Furthermore, ~\cite{Roberg2018} studied this process in
dependence on the plasma beta parameter. They recognized two regimes of this
process: a) the regime with the double layer for low values of the plasma beta
parameter, and b) the regime with the whistlers for the high beta parameter.
Note that in all these studies, Maxwellian or bi-Maxwellian distributions of
particles were considered.

In the paper by~\cite{Kasparova09}, based on fitting of the X-ray spectra of
the coronal flare sources, it was shown that electrons in these sources can be
described by the kappa distributions. This finding was confirmed in the paper
by~\cite{Oka2013,2017ApJ...835..124E,2019ApJ...872..204B}. The Kappa
distribution in flaring regions is also theoretically supported.
\citet{2007PhPl...14j0701R} showed using nonlinear Vlasov and Particle-in-cell
simulations that Langmuir turbulence leads to formation of kappa velocity
distribution.
\citet{2011PhPl...18l2303Y,2012PhPl...19a2304Y,2012PhPl...19e2301Y} in the
series of papers also analytically calculated that the kappa distribution can
be rigorous steady-state solution of the Langmuir turbulence. However, in our
case, the flare coronal sources (plasmoids) are transient phenomena, where the
distribution is generated by the acceleration processes in the current sheet,
where the plasmoids are formed. In the present paper, we try to answer how the
hot plasma with the kappa distribution of electrons is confined in these
coronal sources.

For these reasons, we study an expansion of the hot plasma with the
kappa electron distribution into the cold one. For the hot plasma we
consider an isotropic distribution function because no information about a
possible anisotropy. We choose ratio between mean speed of hot electrons and
thermal velocity of cold ones as $v_\mathrm{he}/v_\mathrm{ce} = \sqrt{10}$.
Such a study is made, according to our knowledge, for the first time. We use a
3-dimensional electromagnetic particle-in-cell (PIC) code, which is
large in one direction and short in other directions. Firstly, for
comparison with the previous studies, we analyze the case, where both hot and
cold plasmas have the Maxwellian distributions (in the following, we call this
case as Maxwell model). Then, we study the case with the hot plasma with the
electron kappa distribution expanding into the cold Maxwellian one (Kappa
model). Finally, the results of both Kappa and Maxwell model are compared and
discussed.

The paper is organized as follows. In Section 2, we describe our numerical PIC
model. The results are in Section 3 and discussion and conclusions in Section
4.

\section{Numerical model}
We use a 3-dimensional electromagnetic PIC code TRISTAN
\citep{1985stan.reptR....B,matsumoto.1993,2008SoPh..247..335K}  with multi-core
Message Passing Interface(MPI) parallelization in domains. The simulation box
in $x$-, $y$- and $z$-directions is $49152 \Delta \times 8 \Delta \times 8
\Delta$, where $\Delta=1$ is the grid size. Thereby, the simulation is
effectively only one-dimensional, but describing all electromagnetic effects.
The simulation box in $x$-direction is divided into two parts: the ``left''
part with hot plasma and the ``right'' part with colder plasma for $x<0$ and
$x>0$, respectively, where $x=0$ ($x$ is in units $\Delta$) corresponds to the
position of $25000 \Delta$ in the numerical box. The length of the ``left'' part is more than a half of the whole simulation box in order to study an expansion for a sufficiently long time.

The simulation time step is $\omega_\mathrm{pe}t = 0.0125$. The electron
cyclotron frequency is $\omega_\mathrm{ce} = 0.1\,\omega_\mathrm{pe}$. The
magnetic field is along $x$- direction. We consider the hydrogen
(electron-proton) plasma. The initial electron density is the same as the
proton density, i.e., $n_\mathrm{0} = 100$. The proton-electron mass ratio is
chosen $m_\mathrm{i}/m_\mathrm{e} = 100$ to speed-up studied processes.

We use the periodic boundary conditions in $y$- and $z$- directions. In the
$x$- direction, the mirror boundary conditions are applied. However, at the
``right'' boundary of the simulation box, the particles that have their
velocities five times greater than the thermal speed of the cold plasma are not
reflected, but removed. The electrons coming from the hot plasma part
that are removed on the right boundary are much less numerous than those
in the cold plasma. Although the removing of these electrons does not
guarantees charge neutralization on the right boundary, these electrons make
only a very localized effect close to the right boundary. They are far away from
the space of studied processes. In comparison
with simulations with all periodic boundaries, computations with these
boundaries extend the effective size of the simulation box with the same
computational demands.

In the initial state, the hot plasma is located at positions at $x<0$ and
cold plasma is at $x>100$. The transition between the hot and cold plasmas is at
$x = 0-100$. The hot plasma consists of electrons with the kappa distribution
\begin{equation}
f_\kappa(\mathbf{v}) = \frac{n_\mathrm{0}}{2\pi (2\kappa v_\mathrm{he}^2)^{3/2}}
\frac{\Gamma(\kappa + 1)}{\Gamma(\kappa - 1/2) \Gamma(3/2)}
\left( 1 + \frac{\mathbf{v}^2}{2\kappa v_\mathrm{he}^2} \right)^{-(\kappa+1)}
\end{equation}
where $\kappa$ is the spectral index, $\Gamma(x)$ is the Gamma function,
$v_\mathrm{he} = \sqrt{(\kappa - 3/2) k_\mathrm{B}T_\mathrm{he}/(\kappa
m_\mathrm{e})}$ is the mean speed of hot electrons, and $k_\mathrm{B}$
is the Boltzmann constant. In the simulation, we apply $\kappa = 2$ to
emphasize the effects of kappa distribution. We note, that the end of
tail of the distribution function ($v > 12\,v_\mathrm{he}$) is not fully
covered due to the finite number of particles. Effectively, the resulting
distribution is more similar to regularized kappa distribution
\citep{2017EL....12050002S} with $\alpha \lessapprox 0.08$. Also note, that
$\kappa \approx 2$ was observed during the early and impulsive phases of the
solar flare \citep{Dzifcakova18}.

The protons in a hot plasma and electrons and protons in cold plasma have Maxwellian
velocity distribution function with corresponding thermal velocities
$v_\mathrm{hi}, v_\mathrm{ce}$, and $v_\mathrm{ci}$. The Maxwell velocity
distribution function is defined as
\begin{equation}
    f_\mathrm{M}(\mathbf{v}) = \frac{1}{(2\pi v_\mathrm{\alpha}^2)^{3/2}} \mathrm{e}^{\frac{\mathbf{v}^2}{2 v_\mathrm{\alpha}}},
\end{equation}
where $v_\mathrm{\alpha} = \sqrt{k_\mathrm{B}T_\alpha/m_\alpha}$ means the
thermal velocity of the particle $\alpha$ with mass $m_\alpha$. The ratio of
the characteristic velocity of the hot plasma and the thermal velocity of the
cold plasma is $v_\mathrm{he}/v_\mathrm{ce} = v_\mathrm{hi}/v_\mathrm{ci} =
\sqrt{10}$. The thermal velocity of the cold component is $v_\mathrm{ce} =
0.006488\,c$, where $c$ is the light speed. That corresponds to the
temperature of cold electrons and protons 250~kK, hot protons 2.5~MK and hot
electrons 10~MK. These temperatures correspond to those in solar flares. The
transition between the hot and cold plasmas is implemented separately for each
species. The protons have Maxwellian distribution in both parts. We implemented
a linear transition in temperature. The implementation of transition from kappa
velocity distribution to Maxwellian is not trivial without generating deformed
velocity distribution. We implemented the transition in both characteristic
velocity and $\kappa$ index. The square of the velocity $v^2$ is scaling
linearly from $v_\mathrm{he}^2$ to $v_\mathrm{ce}^2$. The $\kappa$ index is
scaling from $\kappa=2$ to $\kappa = \infty$ (Maxwell distribution). We
implemented the linear scaling of its reciprocal value $1/\kappa$ in the
interval $1/2 - 1/\infty = 0.5 - 0$.

The Debye length $\lambda_\mathrm{c} = v_\mathrm{ce}/\omega_\mathrm{pe}$ is
$0.260\,\Delta$ for the cold part. The hot electron Debye length
$\lambda_\mathrm{h} = v_\mathrm{he} / \omega_\mathrm{pe} \sqrt{(2\kappa -
3)/(2\kappa - 1)}$ equals to $0.474\,\Delta$. The plasma beta parameter, $\beta =
\frac{1}{2} (\omega_\mathrm{pe} v_\mathrm{t} / \omega_\mathrm{ce} c)^2$, is
$\beta = 2.1\times 10^{-3}$ for cold part and $\beta = 2.1\times 10^{-2}$ for
hot part.

The time of presence of hot electrons in our numerical box can be estimated as
$2L/v_\mathrm{tail} \approx 6.6 \times 10^5 $~time steps $ =
8250\,\omega_\mathrm{pe}t$, where the length of the hot part is
$L=25000\,\Delta$. $v_\mathrm{tail} = 0.15\,c$ is the typical speed of
the generated electrons in the tail of the distribution function if we take
into account that the numerical particle density is limited by a finite number
of numerical particles. The factor of 2 corresponds to the propagation of hot
tail electrons in the simulation box. One group of hot tail electrons
propagates directly to the right direction. Other hot tail electrons have the
initial velocity to the left direction, and then they are reflected to the
right by the left mirror boundary, thereby also contributing to the electron
flux at the hot-cold transition region. On the other hand, our simulations last
for $3\times10^5$~time steps~$ = 3750\,\omega_\mathrm{pe}t$. Thus, at the end
of our simulations, approximately half of all hot tail electrons still remain
in the hot part of the simulation box. If we assume the plasma
density at the thermal front in the flare loop as $n_e = 10^{10}$~cm$^{-3}$
\citep{1997ApJ...480..825A}, then the simulation time corresponds to
4.2~$\mu$s.

For comparison, we also performed a simulation that has all parameters the
same, except it contains hot electrons with Maxwellian velocity distribution.
The ratios $v_\mathrm{he}/v_\mathrm{ce} = v_\mathrm{hi}/v_\mathrm{ci} =
\sqrt{10}$ remains same. The hot electron Debye length is $0.82\,\Delta$. The
simulation with the kappa distribution of hot particles we designate as ``Kappa
model'', and the simulation with the Maxwellian distributions only as ``Maxwell
model''.

\section{Results}
\subsection{Maxwell model}

\begin{figure}
    \centering
    \gridline{
        \fig{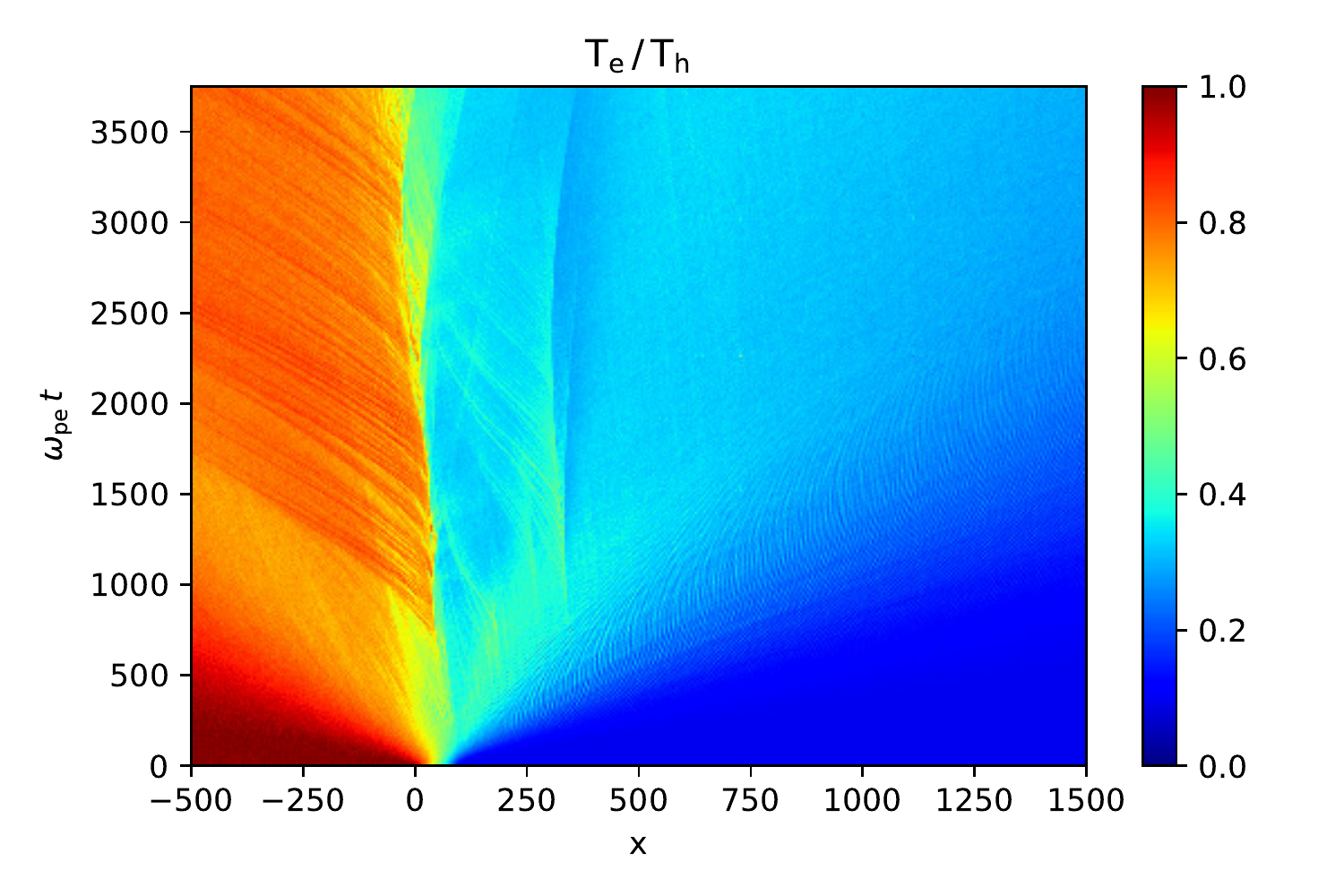}{0.49\textwidth}{(a)}
    }
    \gridline{
        \fig{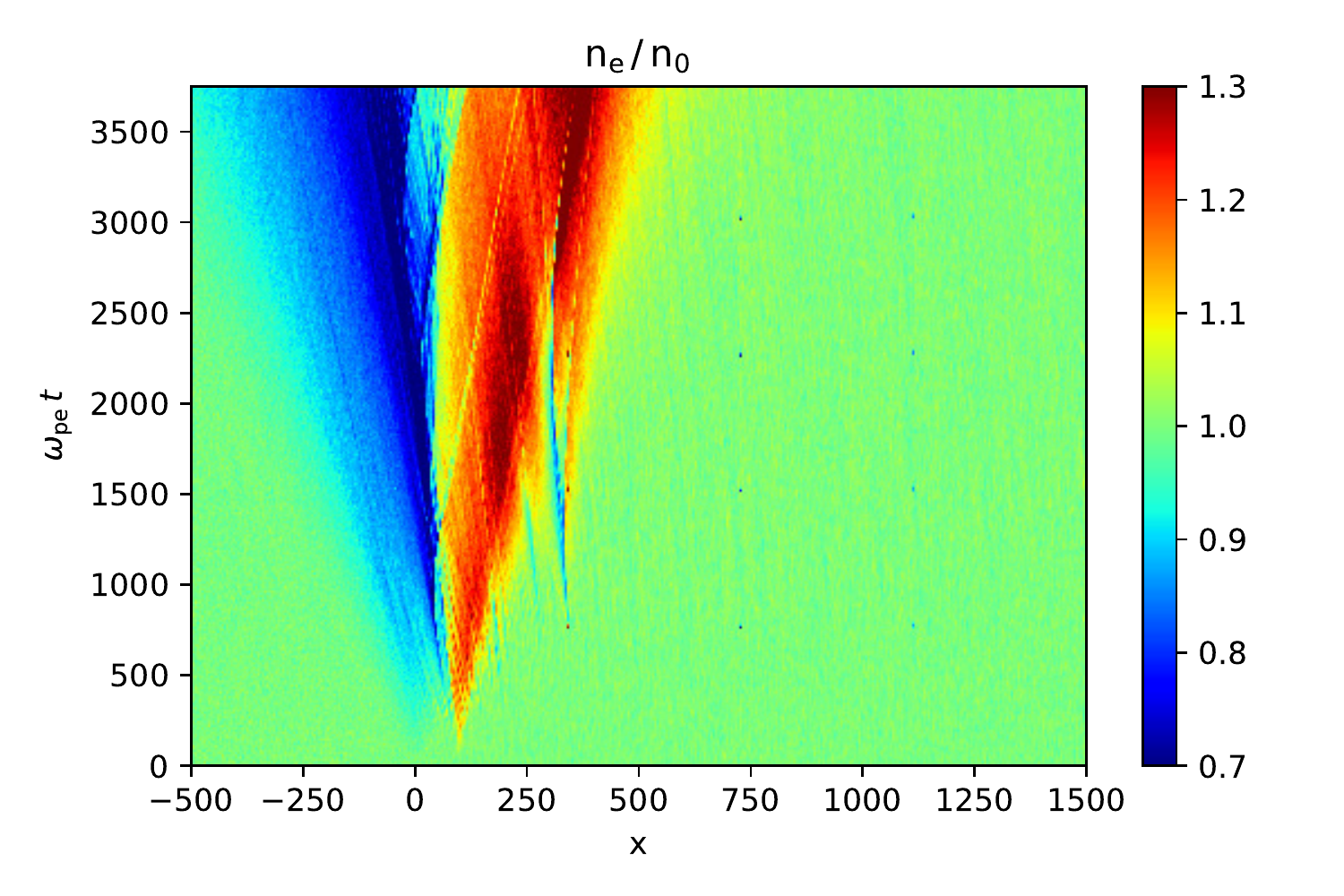}{0.49\textwidth}{(b)}
    }
    \gridline{
        \fig{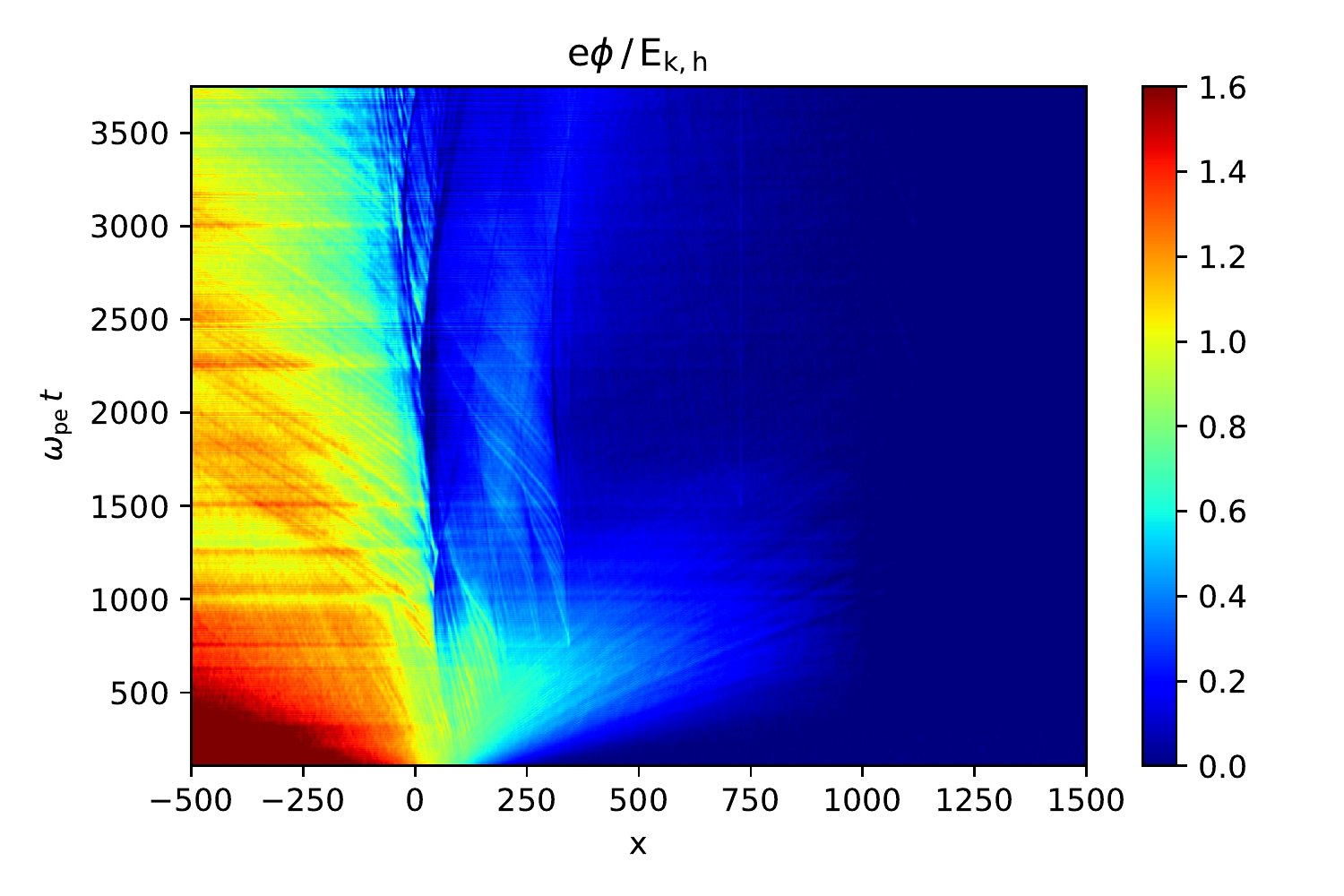}{0.49\textwidth}{(c)}
    }
    \caption{Maxwell model:
        (a) Evolution of the electron temperature $T_e$ normalized to the initial hot plasma temperature $T_h$.
        (b) Evolution of the electron density $n_e$ normalized to the initial particle density $n_0$.
        (c) Electric potential.}
    \label{fig1}
\end{figure}

Figure~\ref{fig1} shows an evolution of the electron temperature, electron
particle density, and electric potential energy. These quantities are taken
along $x-$ axis and always averaged in $y-$ and $z-$ directions. The
temperature of species $\alpha$ is computed as $T_\alpha k_\mathrm{B} = m_\alpha \langle (
v_\alpha - \langle v_\alpha \rangle )^2 \rangle$, where the mean value is
computed over all particles inside a grid cell. The temperature and density are
normalized to the initial hot plasma temperature and the initial density $n_0$,
respectively. The electric potential is expressed here and in the following as
the potential energy normalized to the mean initial kinetic energy of hot
electrons $E_\mathrm{k,h} = \frac{1}{2} m_\mathrm{e} v_\mathrm{he}^2$ and set
to zero at $x$ = 1000. It is also smoothed along time interval
$18.6\,\omega_\mathrm{pe} t$ (1500~time steps).

\begin{figure}
    \centering
    \includegraphics[width=0.4\textwidth]{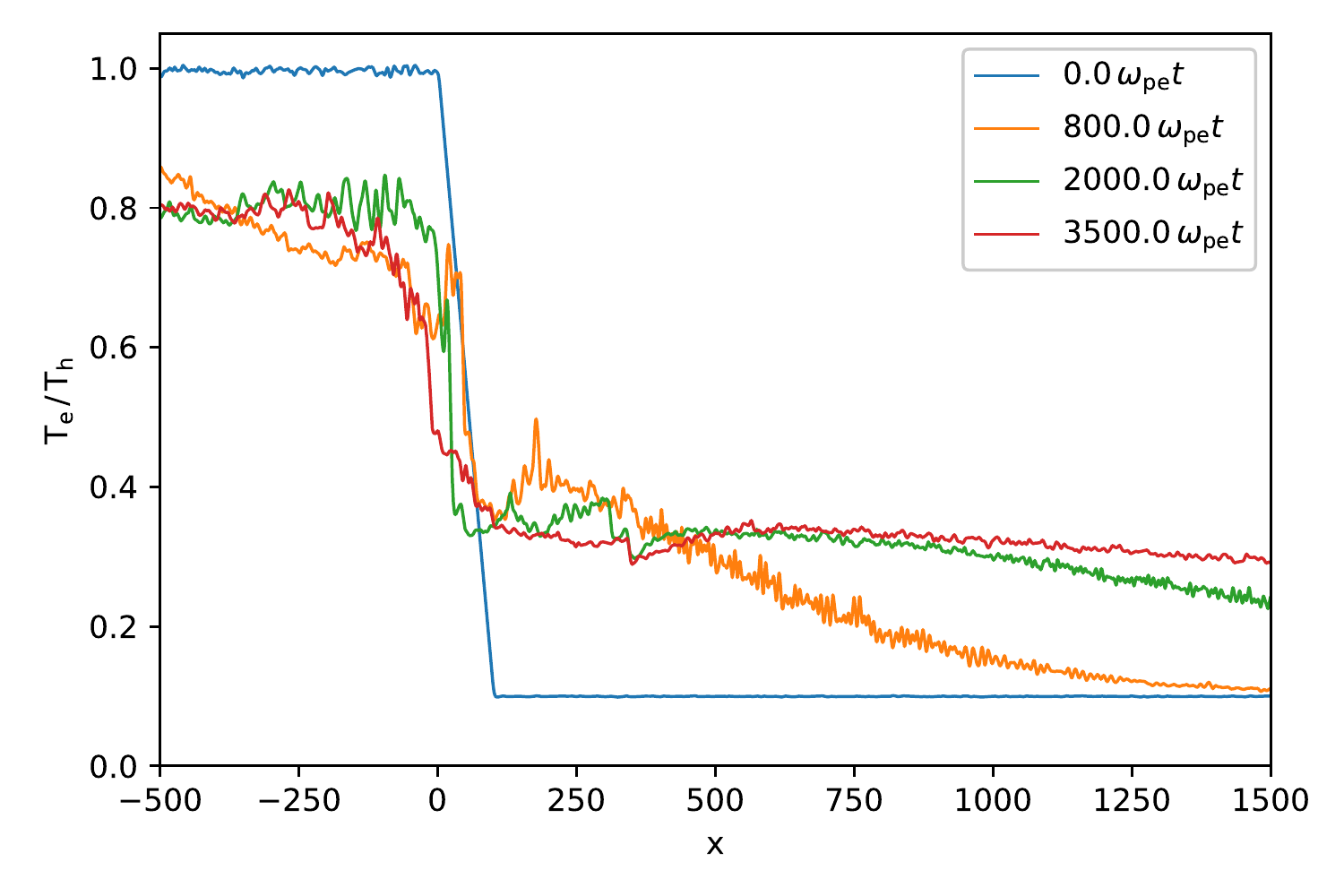}
    \caption{Maxwell model: Electron temperature profiles at four times. The
        temperature is normalized to the initial temperature of the hot plasma.}
    \label{fig2}
\end{figure}

As seen in Figure~\ref{fig1}a, at the hot-cold plasma transition region and
starting from the initial time, the temperature of the hot plasma decreases,
and that of the cold plasma increases. It is owing to a free-streaming of hot
plasma electrons into the cold plasma. It is associated with a decrease of the
electron plasma density in the hot plasma part ($x < 0$) and density increase
in the cold plasma part ($x > 0$) (Figure~\ref{fig1}b). Simultaneously,  at the
location close to $x \sim 0$, the electric potential jump is formed and
drifting with the velocity $8\times 10^{-4}\,c$ to the region with $x < 0$
(Figure~\ref{fig1}c). This velocity agrees to the local ion-acoustic
speed. The potential jump corresponds to the sharp decrease in the electron
temperature profiles as shown in Figure~\ref{fig2}, which indicates a
suppression of the hot electron flux from the hot plasma part into the colder
one.

\begin{figure}
    \centering
    \gridline{
        \fig{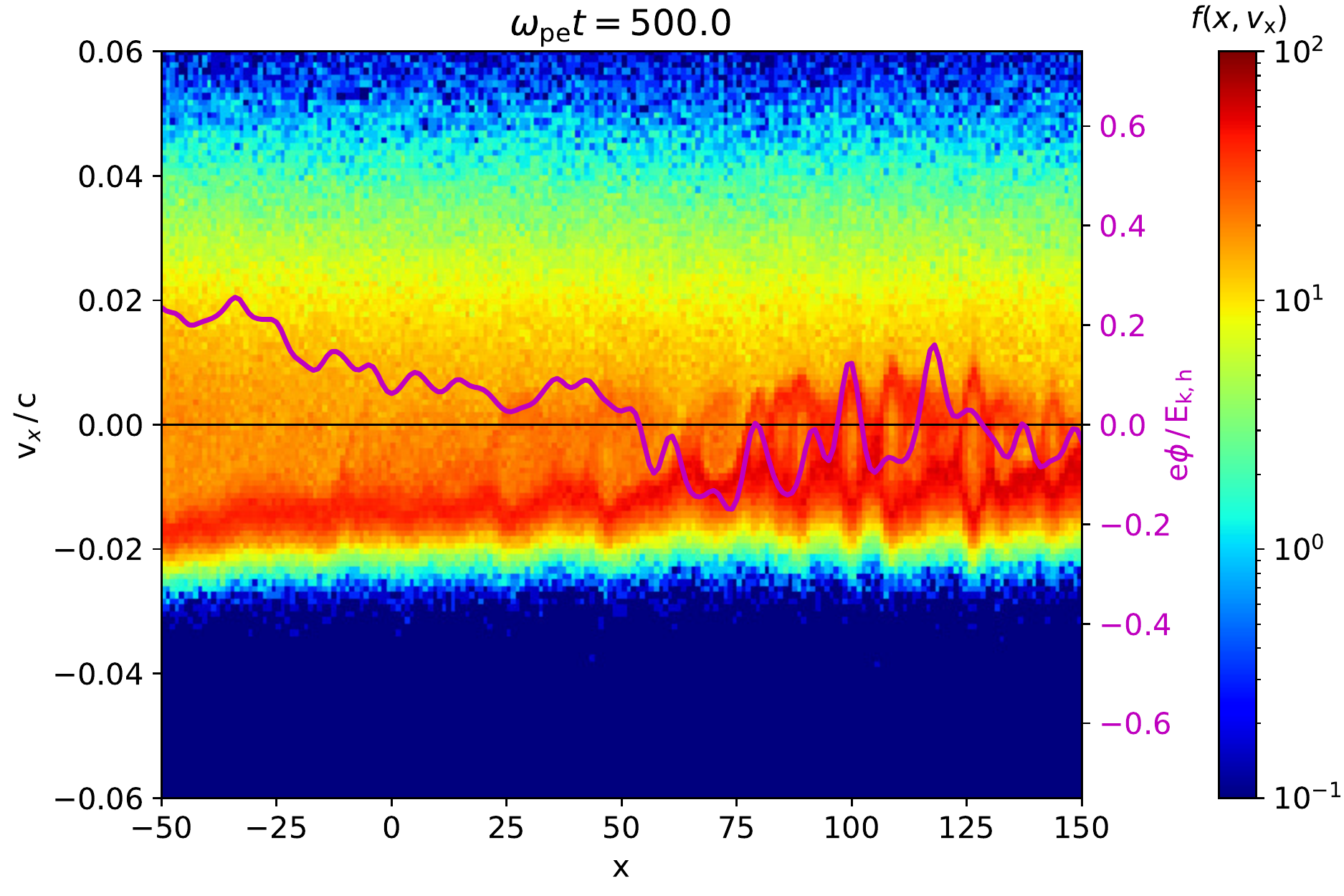}{0.49\textwidth}{(a)}
    }
    \gridline{
        \fig{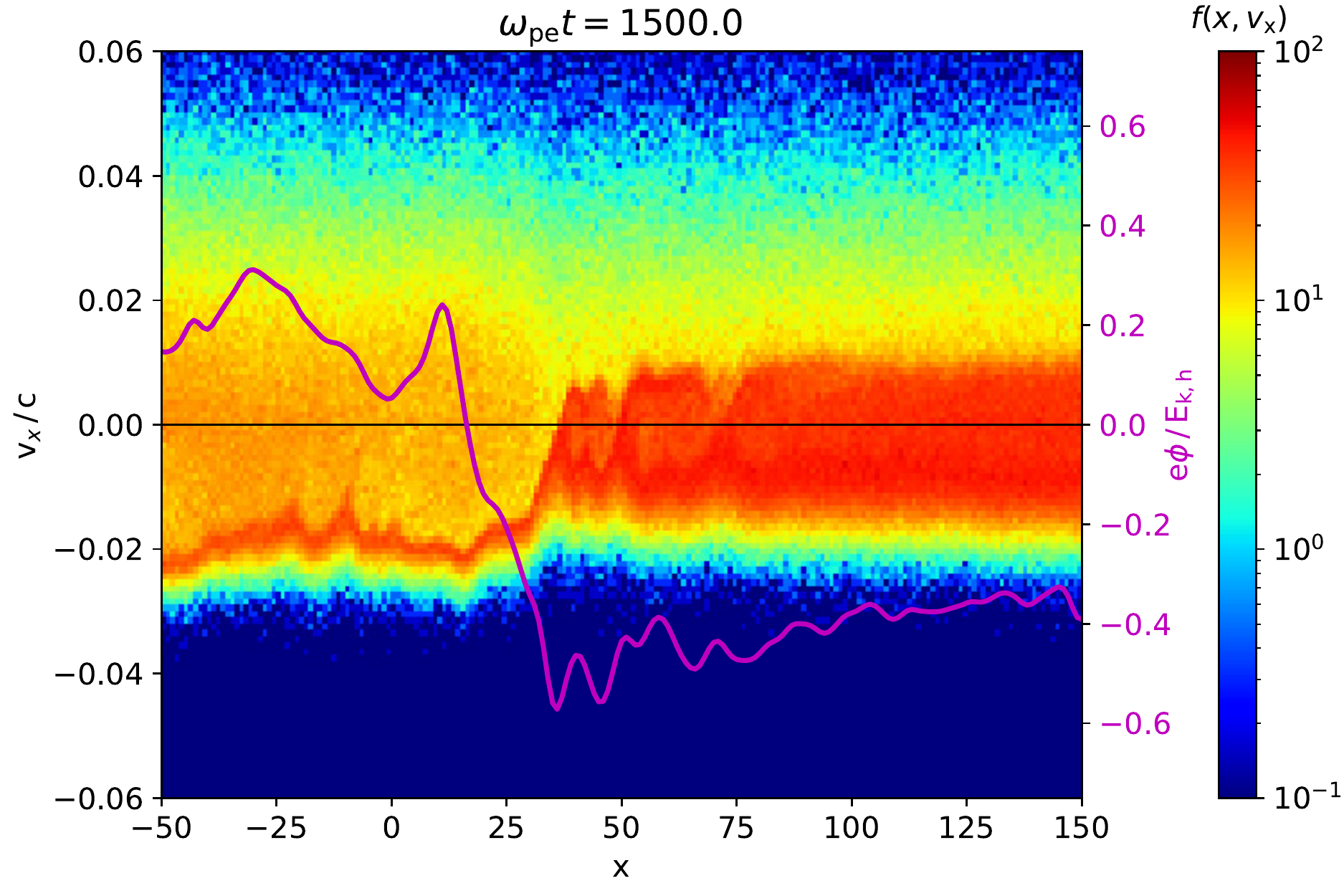}{0.49\textwidth}{(b)}
    }
    \gridline{
        \fig{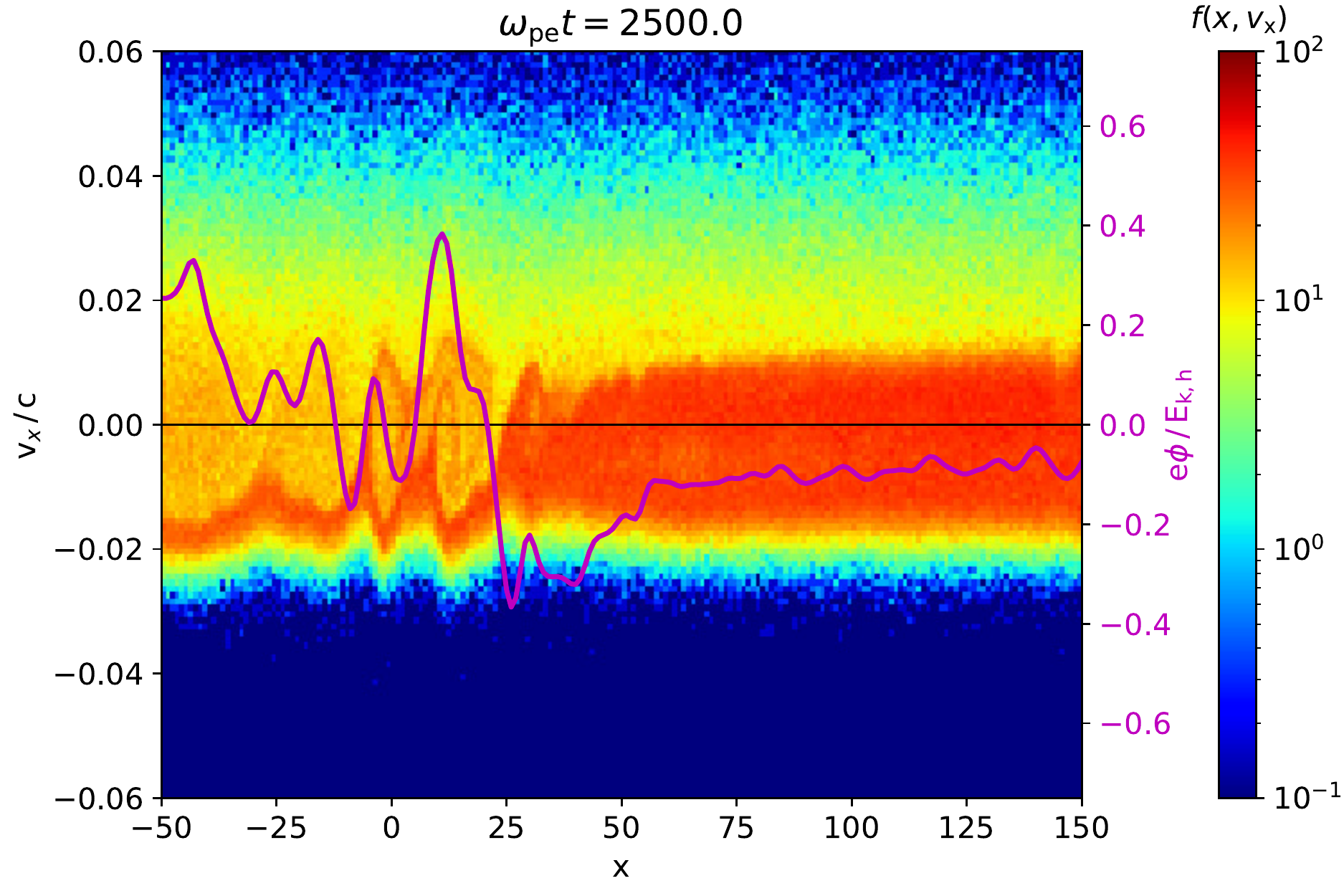}{0.49\textwidth}{(c)}
    }
    \caption{Maxwell model: The electron velocity $v_x$ distribution
    along $x-$axis with the electric potential (magenta line) overlaid.
    (a) 500~$\omega_\mathrm{pe}t$, (b) 1500~$\omega_\mathrm{pe}t$, and
    (c) 2500~$\omega_\mathrm{pe}t$. The distribution color scale is logarithmic.}
    \label{fig3}
\end{figure}

To see more details about these processes, we show the
electron velocity $v_x$ distribution together with the electric potential at
three times $\omega_\mathrm{pe}t = $ 500, 1500, and 2500 in Figure~\ref{fig3}.
As seen here, in
all these times, the electrons from the hot plasma region are streaming with the
positive velocities to the cold plasma region. At time
500$\,\omega_\mathrm{pe} t$ and for  $x<70$ the maximum of the distribution is
shifted to the negative value $v_\mathrm{x} = -0.015\,c$, thus forming the
return current, which compensates the current of streaming hot electrons.
Namely, the total electric current needs to be close to zero. In accordance
with the description of these processes in \cite{Li2012}, the return current
generates the ion-acoustic waves by the Buneman instability.  At positions
$x>70$, the distribution is disturbed, even multi-peaked. The electric
potential at this time is waved, but not still forming a significant jump.

It happens in later time. At 1500$\,\omega_\mathrm{pe}$, the thermal front is
fully developed, and the electric potential has a form, which is typical for the
double-layer (DL). The DL restrains all electrons with kinetic energy lower
than $\approx 0.8\,E_\mathrm{k,h}$ on its left side. The electrons with higher
velocity pass, and their kinetic energy is decreased by the potential jump
(cooling). The electrons flying from right to left are not confined, and they
gain some energy passing the DL (heating).

In the following times, the DL evolves. For example, at
2500$\,\omega_\mathrm{pe}$ it is partly deformed. Especially its left part,
where the potential is varying and forming potential wells. Concurrently, the
hot electrons are not so strongly confined. They can escape more easily from
the left plasma part thereby reinforcing the return current.

\begin{figure*}
    \centering
    \includegraphics[width=0.85\textwidth]{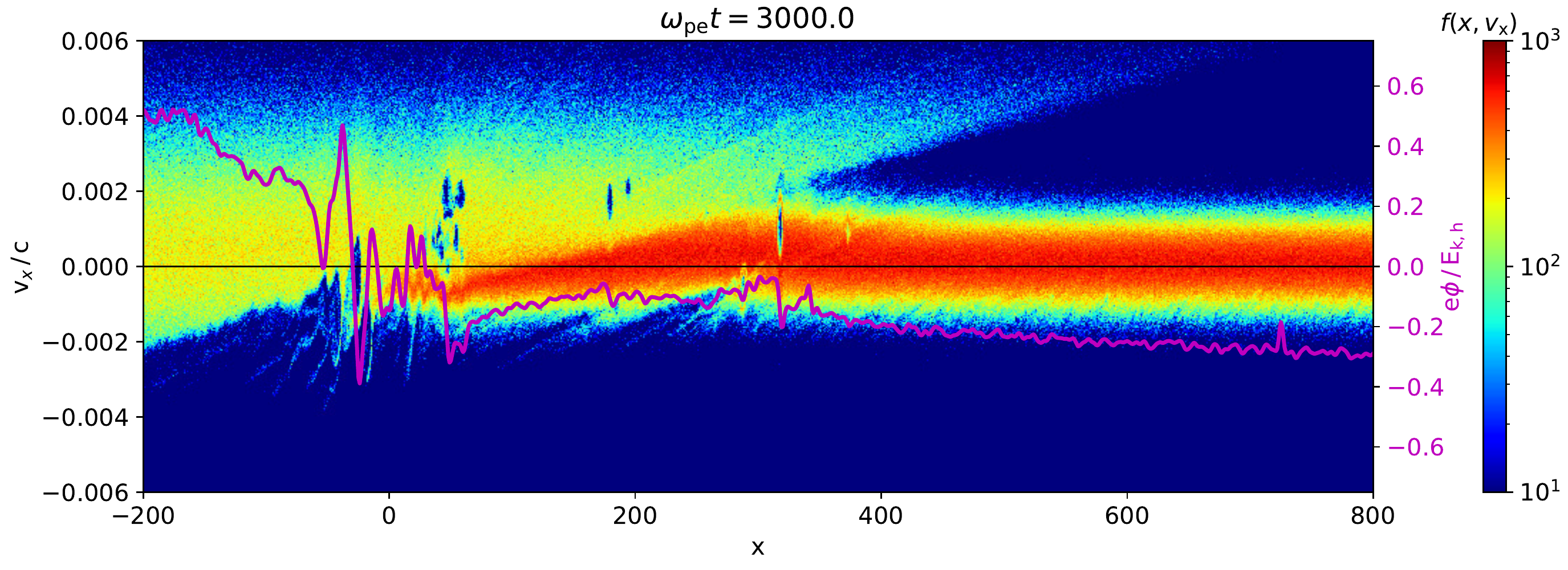}
    \caption{Maxwell model: The proton velocity $v_x$ distribution along
        $x-$axis at $\omega_\mathrm{pe}t = $3~000 with the electric potential
        (magenta line) overlaid.}
    \label{fig4}
\end{figure*}

Furthermore, in Figure~\ref{fig4} we show the proton velocity $v_x$
distribution at 3000~$\omega_\mathrm{pe}t$, i.e., when the DL is slightly
dissipated. The hot protons that were in the initial state in the hot plasma
region form a beam in $x = 350-800$ and have the velocity
$v_\mathrm{x}>0.0025\,c$. The potential jump corresponding to the DL is at
$x=-40$. On the right side of the DL, the protons that are flying to the left
towards the DL are reflected back to the right. The protons that flying from
the left to right pass the DL, and thus, they support the proton beam. In the
system, there are also small DLs at $x \sim 50$, and $x \sim 320$. The potential
increases between them. These DLs influence the proton distribution.

In summary: The results of our Maxwell model are similar to those shown in the
papers by ~\cite{Li2012,Sun19}. Some small differences are owing to that all
our distributions are taken as isotropic and in the hot plasma part there are
hot protons. The results from this section will be used for comparison with
those in the Kappa model below.

\subsection{Kappa model}

\begin{figure}
    \centering
    \gridline{
        \fig{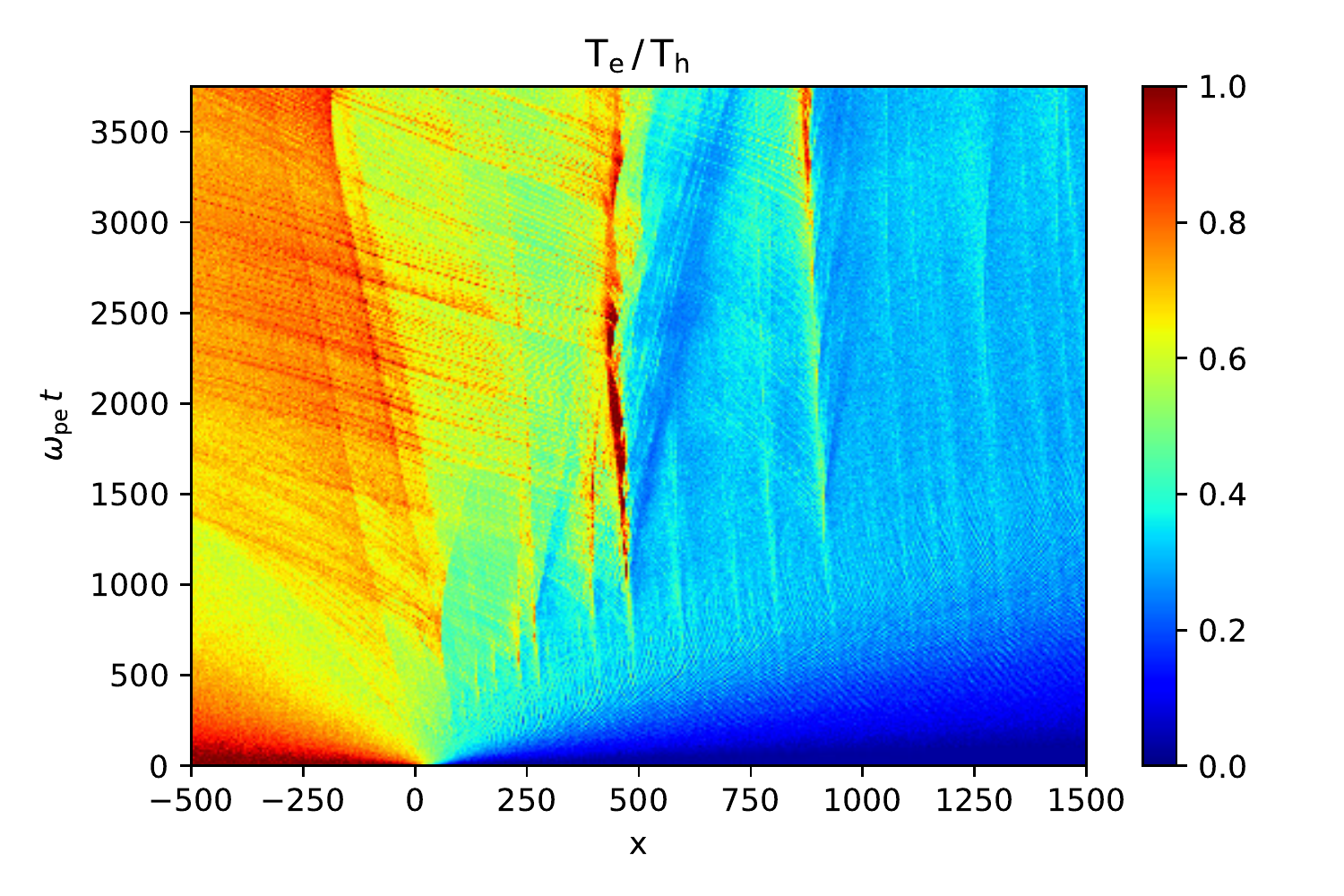}{0.49\textwidth}{(a)}
    }
    \gridline{
        \fig{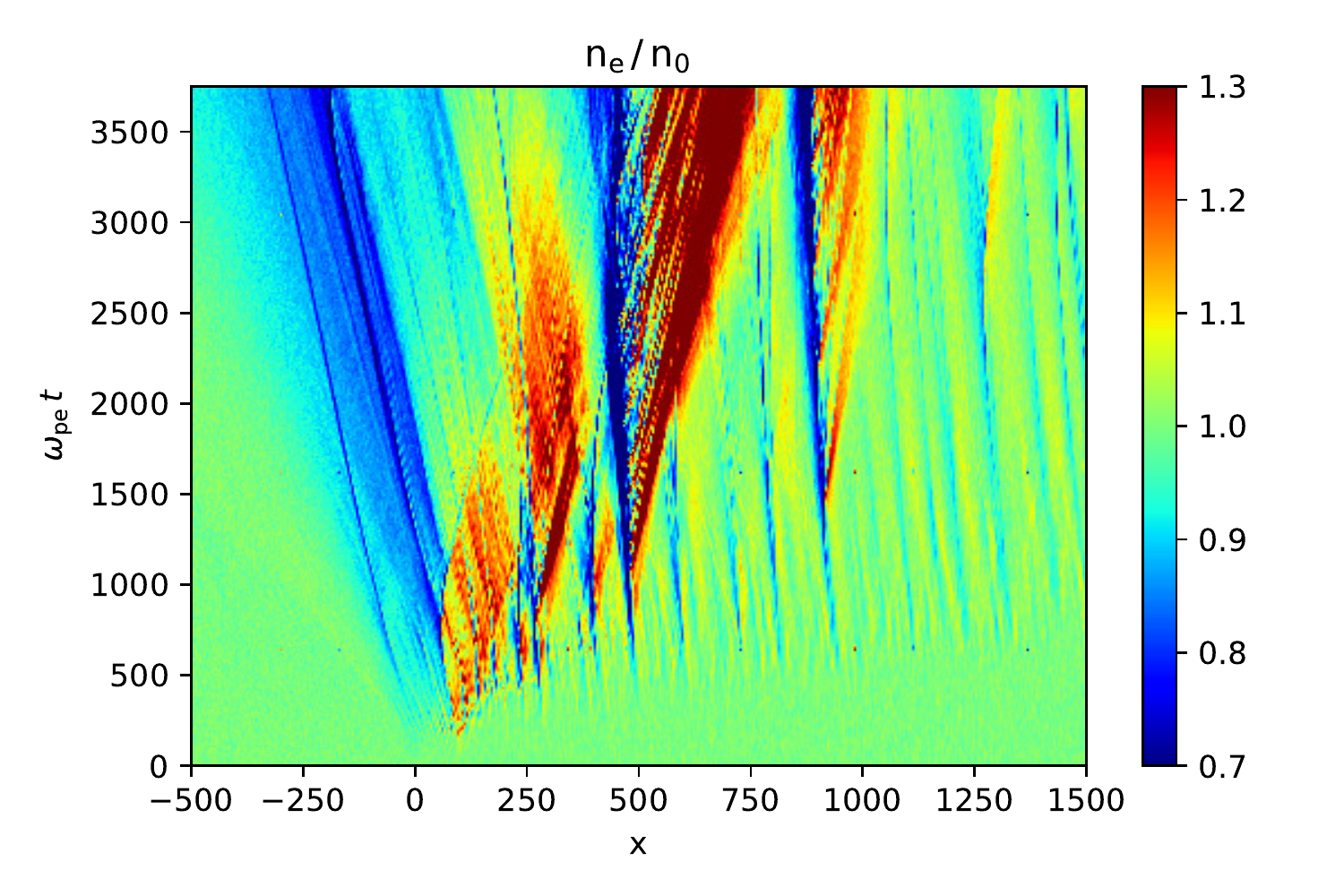}{0.49\textwidth}{(b)}
    }
    \gridline{
        \fig{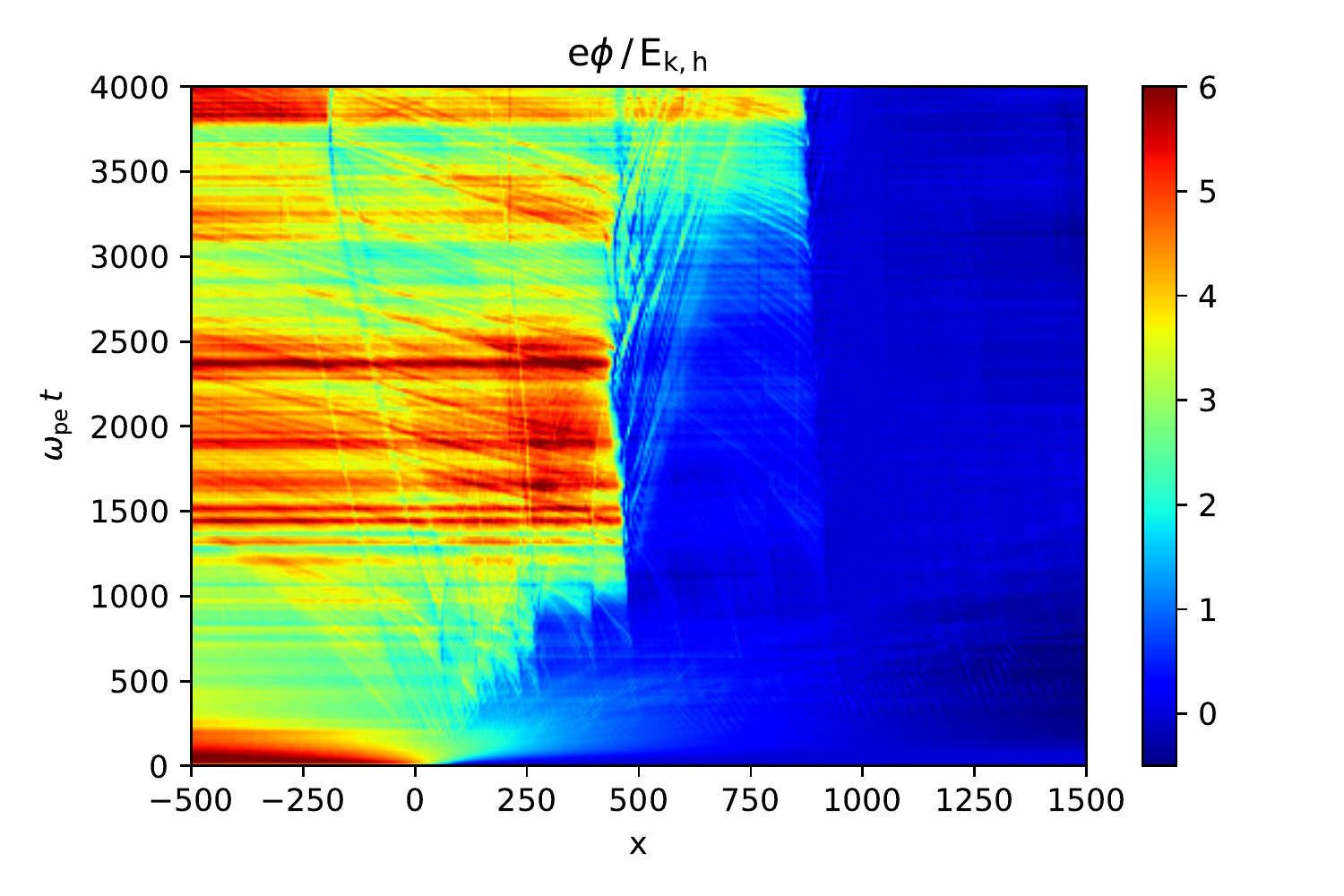}{0.49\textwidth}{(c)}
    }
    \caption{Kappa model:
        (a) Evolution of the electron temperature $T_e$ normalized to the initial hot plasma temperature $T_h$.
        (b) Evolution of the electron density $n_e$ normalized to the initial particle density $n_0$.
        (c) Electric potential. Compare with Figure~\ref{fig1}.}
    \label{fig10}
\end{figure}

The evolution of the electron temperature, plasma density, and electric
potential in Kappa model is shown in Figure~\ref{fig10}. Contrary to the Maxwell
model, the Kappa model shows a multi-front solution. Firstly, not very
distinctive front (Front 1)  is formed at $x\sim -50$ and $x \sim 50$
(Figure~\ref{fig10}a). The other fronts are formed on the right side of it in a
disturbed plasma in the cold plasma region. At 1000~$\omega_\mathrm{pe}
t$ and $x \sim 500$ a new distinct front is formed  (Front 2). Its temperature
is increasing until 2000~$\omega_\mathrm{pe}t$ when it reaches a maximum.
Then, it slowly dissipates. Shortly after a formation of this front, the new
front (Front 3) is generated at $x \sim 900$. From the beginning, the front is
weak, but at times after 3000~$\omega_\mathrm{pe}t$, its temperature raises.

Shortly after the start of the simulation, density waves are created.
Because the plasma conserves the electric neutrality, almost the same waves are
created in the proton and electron density.
The waves are reinforced during the evolution,
and their edges mutually intersect. The most distinct density depressions are
connected with Front 1, Front 2, and Front 3 (Figure~\ref{fig10}b). Between
these main density depressions, there are feeble ones that gradually appear and
dissipate.

At the location of Front 1, there is only some potential well, not the DL as in
Maxwell model. On the other hand, a strong DL appears at
1000~$\omega_\mathrm{pe} t$ and $x=485$ in connection with Front 2. Its
potential jump is about $e\phi = 6\,E_\mathrm{k,h}$. In the time interval
2500--3000$\,\omega_\mathrm{pe} t$, this DL is disintegrating. The hot electrons
and cold protons, that have been detained by this DL, escape. Since
3000~$\,\omega_\mathrm{pe} t$, the DL reinforces at the position of Front 3.

\begin{figure}
    \centering
    \includegraphics[width=0.49\textwidth]{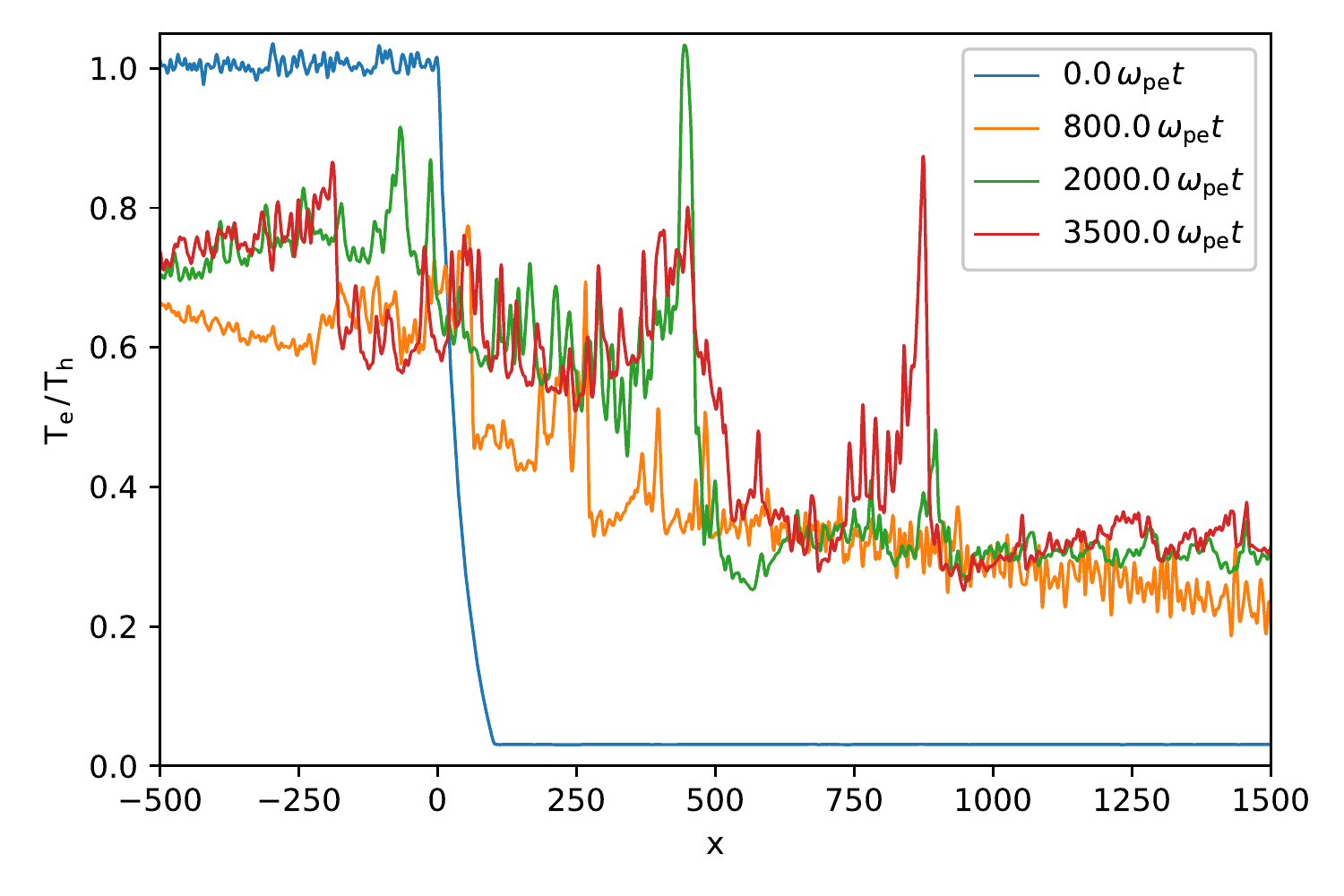}
    \caption{Kappa model: Electron temperature profiles at four times. The temperature is
    normalized to the initial temperature of the hot plasma component.
    Compare with Figure~\ref{fig2}.}
    \label{fig11}
\end{figure}

Figure~\ref{fig11} presents a detailed view of temperate profiles at four
times. At time 800~$\omega_\mathrm{pe} t$ the temperature creates transition
between the hot and cold parts of the model as caused by a free-streaming
electrons from hot to cold plasma. The first, but transient, enhancements are
in $x \sim 50$, $x \sim 250$, and $x \sim 400$. Only the enhancement at $x \sim
50$ (corresponding to Front 1) sustains, and it is slowly moving to the left.
At 2000~$\omega_\mathrm{pe} t$, the highest temperature enhancement is at the
location of Front 2. Front 3 is here illustrated by the temperature enhancement
at 3000~$\omega_\mathrm{pe}$. In comparison with the thermal front in the
Maxwell model, expressed as the one-side temperature step, Front 2 and
3 can be better described as temperature enhancements. Their temperatures are
lower at both front sides. On the left side the temperatures has a much
smoother decrease than on the right side.

\begin{figure*}
    \centering
    \gridline{
        \fig{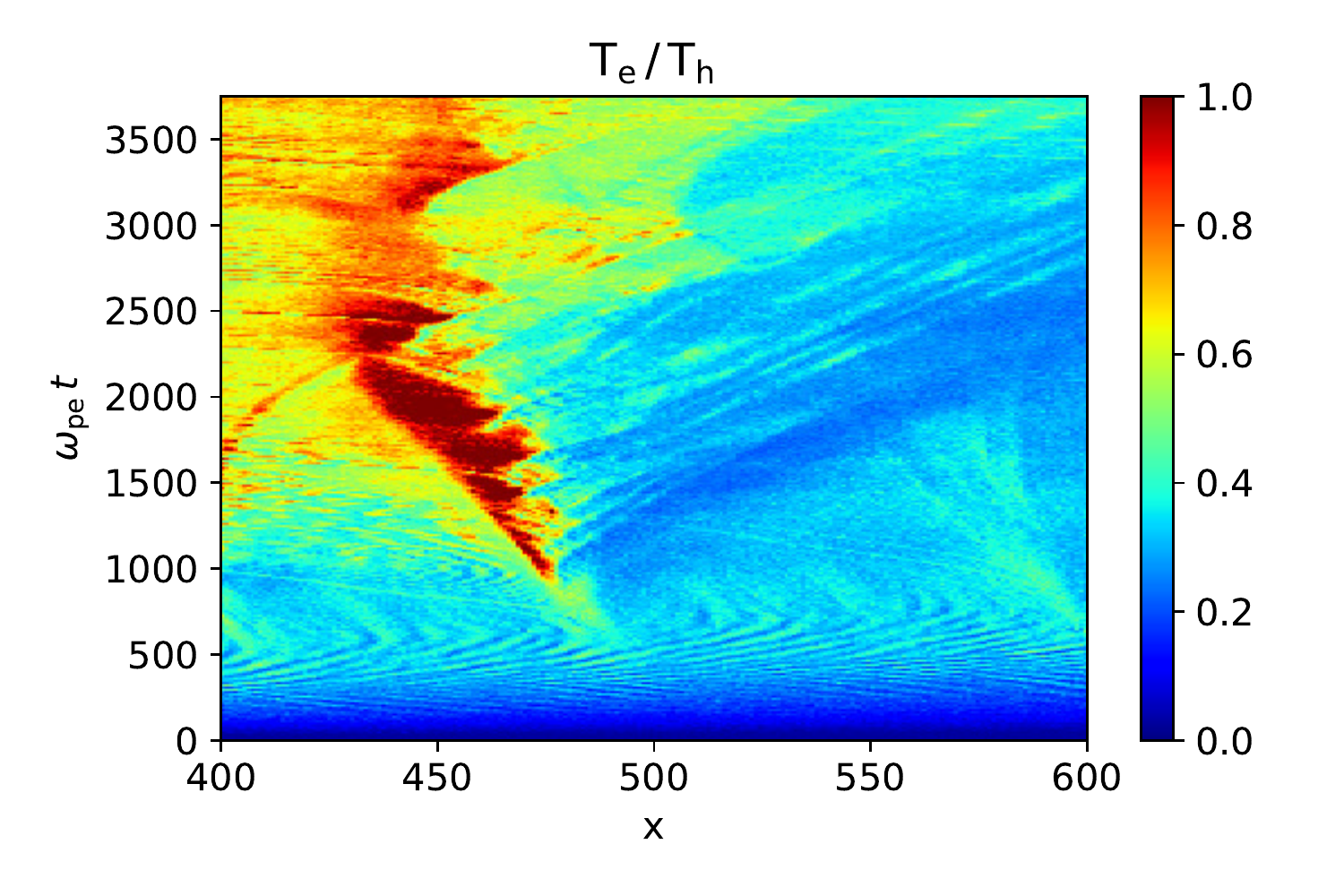}{0.49\textwidth}{(a)}
        \fig{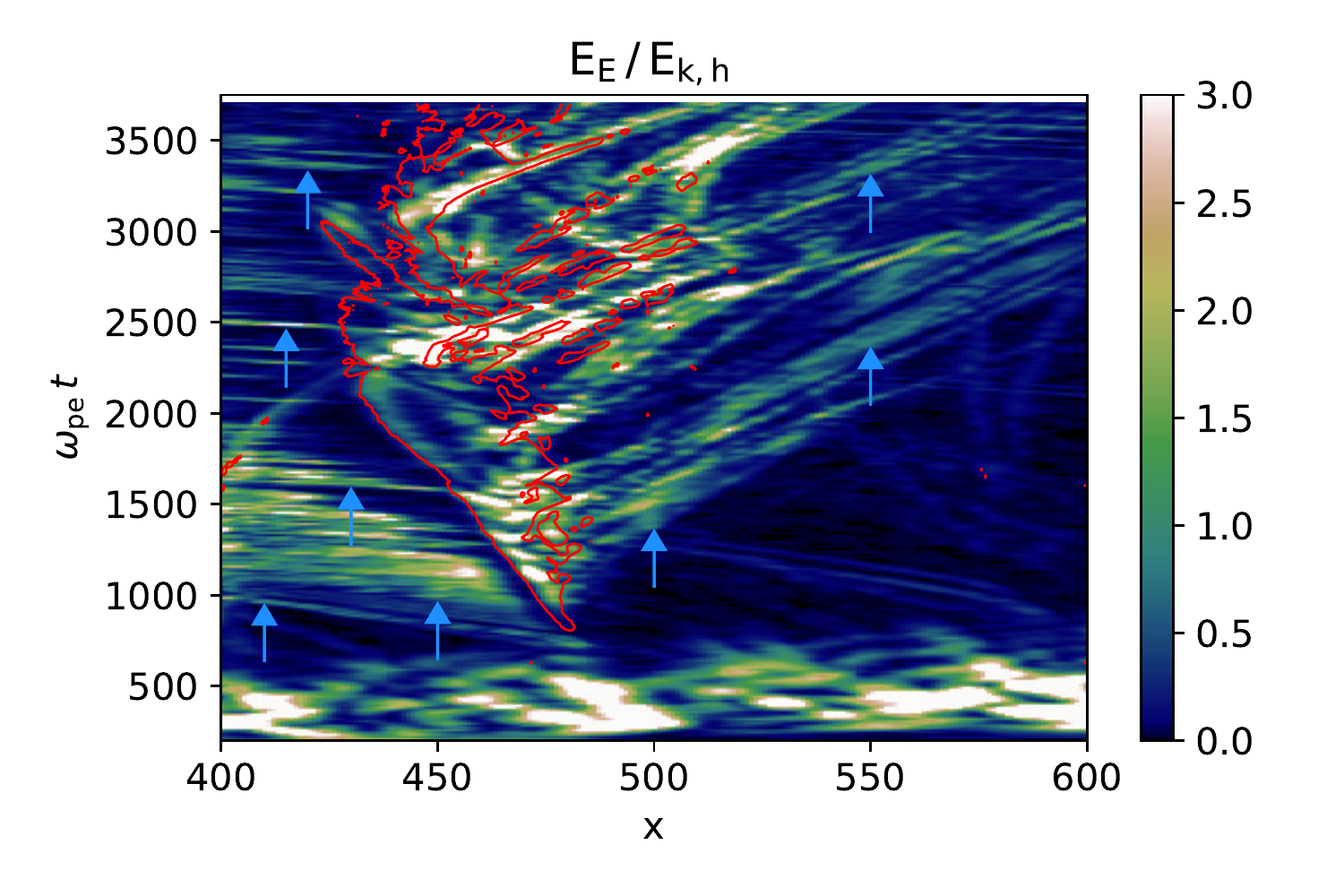}{0.49\textwidth}{(c)}
    }
    \gridline{
        \fig{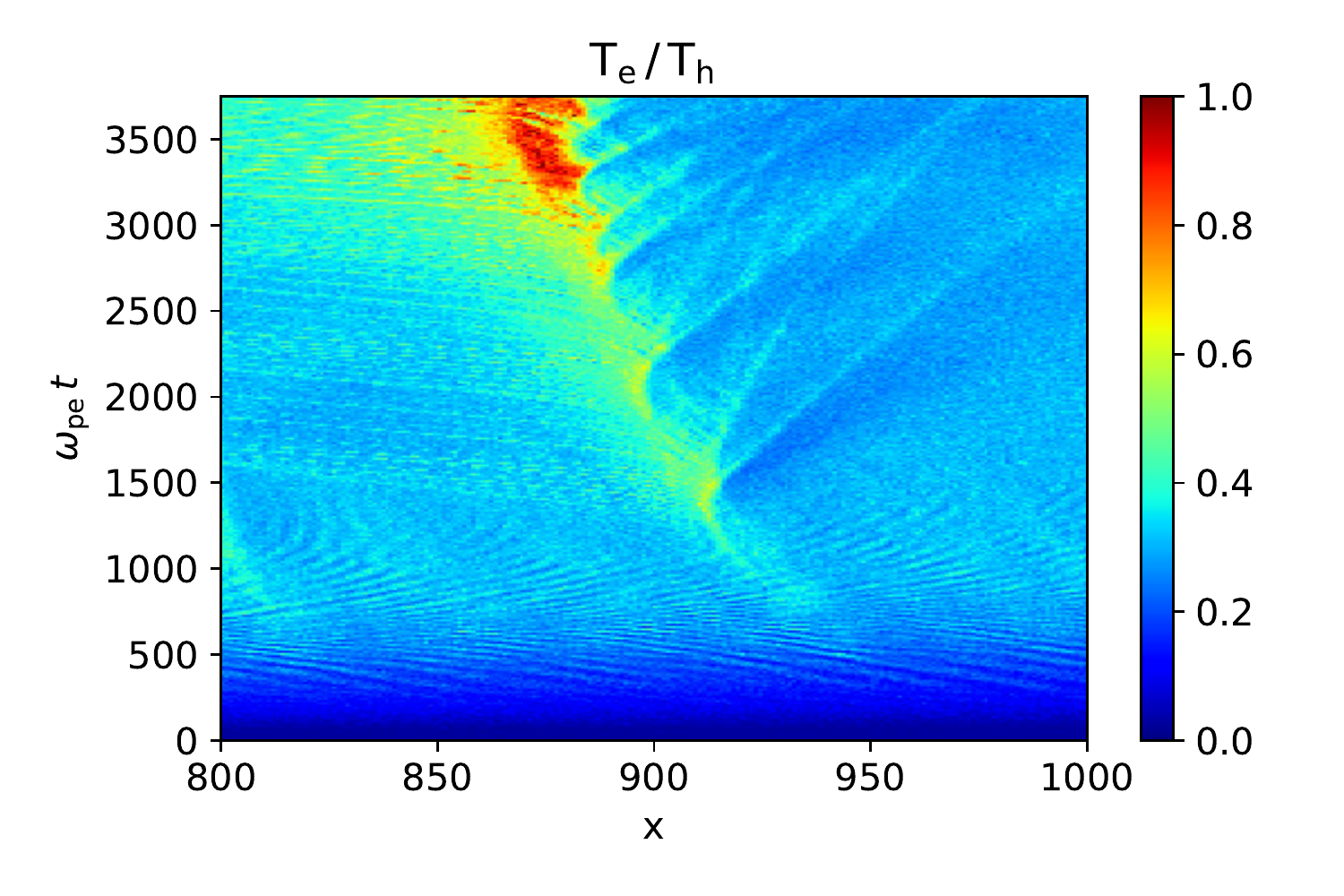}{0.49\textwidth}{(b)}
        \fig{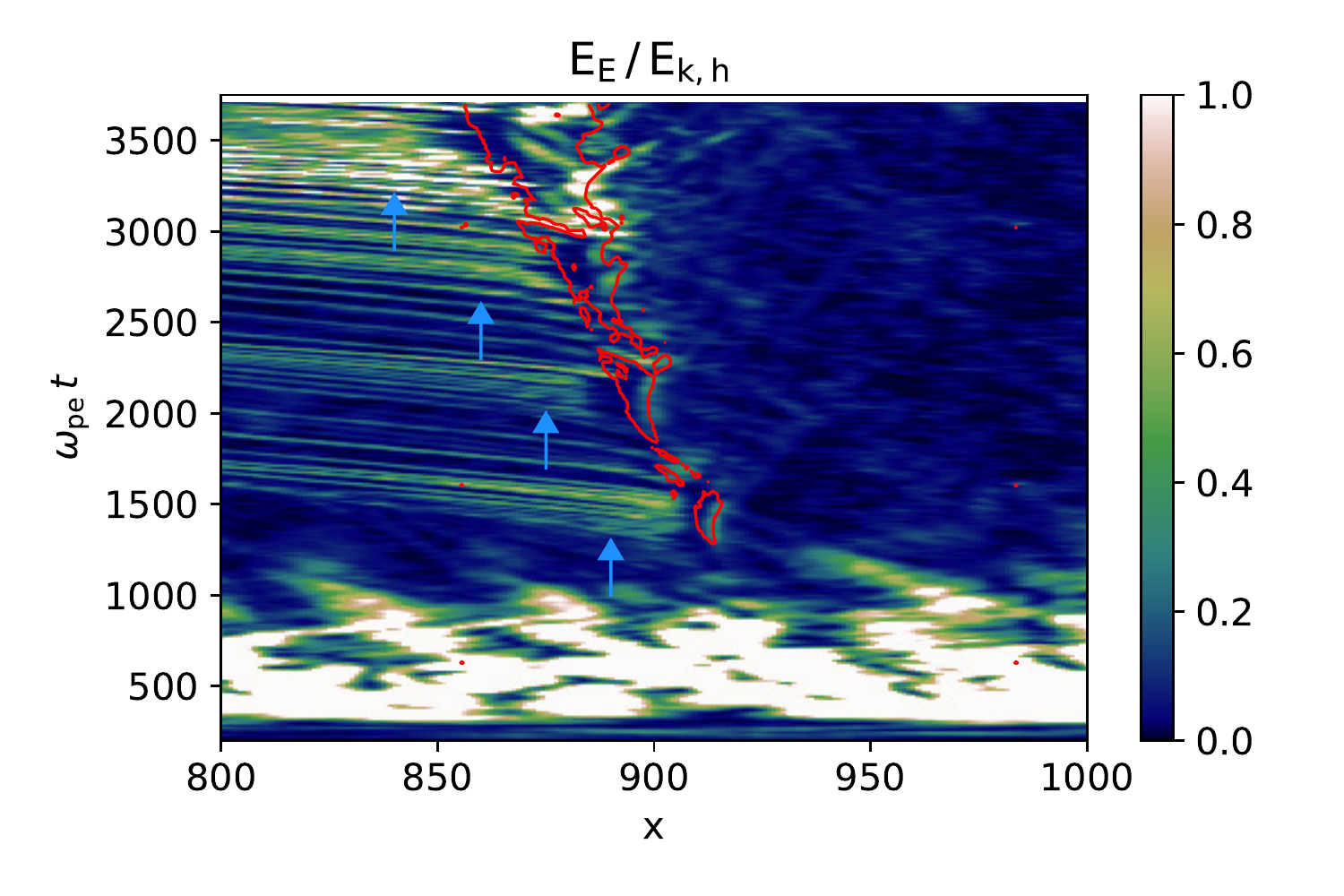}{0.49\textwidth}{(d)}
    }
    \caption{Kappa model: Detailed view on the temperature evolution of Front 2 (a) and Front 3 (b) and the corresponding electric field energy ((c)
    and (d)). The electric field energy density is normalized to the kinetic energy of hot
electrons $E_\mathrm{k,h}$. It is overlaid by contours of the electron density
with $0.7\,n_\mathrm{0}$ (red line). Selected individual Langmuir wave
packets are denoted by blue arrows.}
    \label{fig12}
\end{figure*}

Detailed view on the electron temperature evolution of Front 2 and 3, as well
as the electric field energy density is shown in Figure~\ref{fig12}. The
temperature enhancements are located at the regions where the density
depressions are; see the red contours in Figure~\ref{fig12}c,d. The
left edge of Front 2 moves to the left with the velocity $1\times 10^{-3}\,c$,
i.e., with the velocity close to local ion-acoustic speed until
2000~$\omega_\mathrm{pe} t$. Then, the front becomes disturbed, and its left
edge is smoothed. The right edge is sharper for the whole time, but its
position changes more rapidly. After the time 2500~$\omega_\mathrm{pe} t$, a
motion of the temperature enhancement turns from the left direction to the
right one. Front 3 moves all the time to the left more slowly than Front 2
because it is surrounded by a colder plasma. Its velocity is about
$-6\times10^{-4}\,c$ and the surrounding plasma ion-acoustic speed is $7\times
10^{-4}\,c$.

The electric field energy in Figure~\ref{fig12}c,d is the energy of the
electrostatic waves. As seen here, there are electrostatic waves at space around Front 2 and 3 everywhere till 500--1000~$\omega_\mathrm{pe} t$ when they are absorbed. Then,
the electrostatic waves appear at the edges of the density depressions, where
the gradient of the density is nonzero. They are on the left edge of Front 2
and the right edge of Front 3. Moreover, they also appear when fronts dissipate
or restore; they manifest the temporal changes in the plasma density.

During the front evolution, the Langmuir wave packets \cite{Vladimirov}) are
generated. Individual Langmuir wave packets are denoted by blue arrows
in Figure~\ref{fig12}. They are identified as local electric field
enhancements. They propagate to the left and right from the fronts and have
lower speed than the ion-acoustic speed. For example, the solitons escaping
from the Front 3 to the left at 1700~$\omega_\mathrm{pe} t$ have their velocity
about $-9\times 10^{-3}\,c$, while the characteristic thermal velocity is
$2.3\times 10^{-2}\,c$. In our case, the solitons travel distance up to
$300\,\Delta$ (1150~$\lambda_{d}$) and some of them live longer then
500~$\omega_\mathrm{pe}t$.

\begin{figure}
    \centering
    \gridline{
        \fig{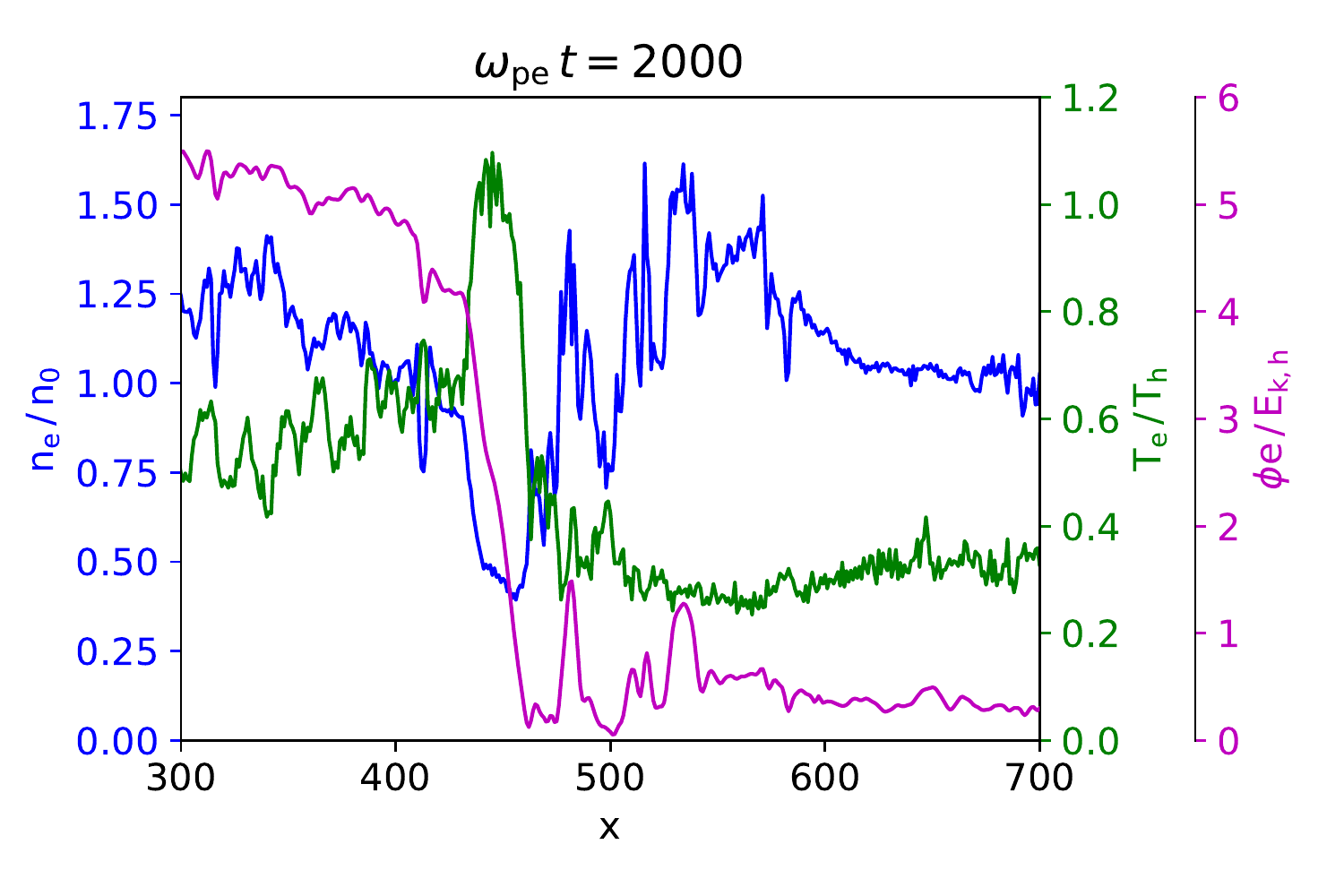}{0.49\textwidth}{(a)}
    }
    \gridline{
        \fig{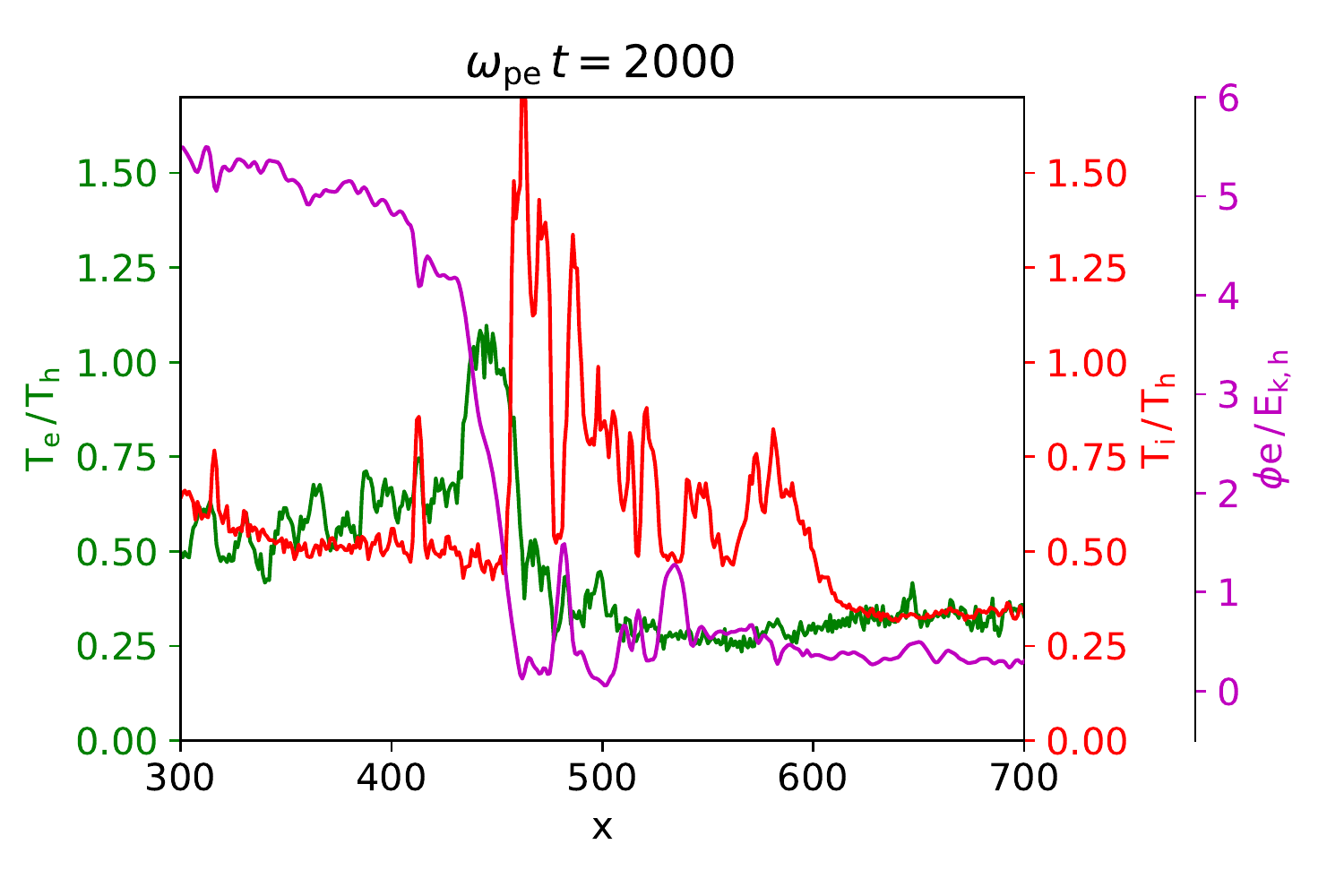}{0.49\textwidth}{(b)}
    }
    \gridline{
        \fig{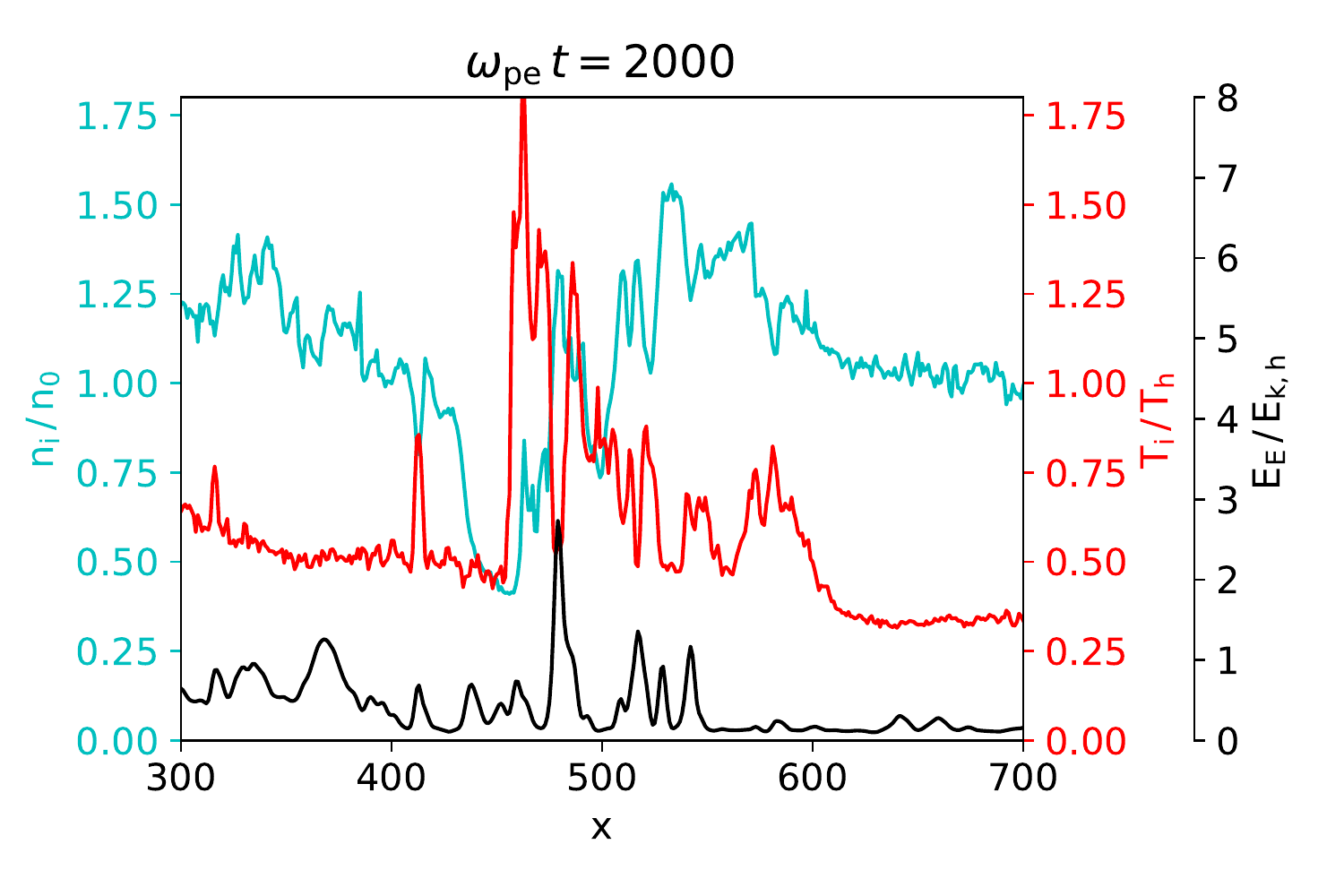}{0.49\textwidth}{(c)}
    }
    \caption{Kappa model: Comparison of temperatures, densities, electrical potential,
        and electric field energy density for Front 2 at $\omega_\mathrm{pe} =$~2000,
    when the front is fully developed. \textit{Blue:} Electron density,
    \textit{Green:} Electron temperature, \textit{Magenta:} Electric potential energy,
    \textit{Red:} Proton temperature, \textit{Cyan:} Proton density,
    \textit{Black:} Electric field energy.}
    \label{fig13}
\end{figure}

Profiles of the temperature, density, electric potential and electric field
energy density for Front 2, at the time of the fully developed front at
$\omega_\mathrm{pe} t = $~2000 are shown in Figure~\ref{fig13}. The electron
temperature peak and the electron and proton density depressions are located at
the same position, where the electric potential steeply decreases (location of
DL). The enhanced proton temperature is shifted a little bit to the right; very
steep on the left side, while on the right side a decrease is much smoother.
See that the maximum of the proton temperature is higher than that of
electrons. The highest peak in the electric energy density is at $x$ =
480 where the densities are enhanced and temperatures have depressions.

\begin{figure*}
    \centering
    \gridline{
        \fig{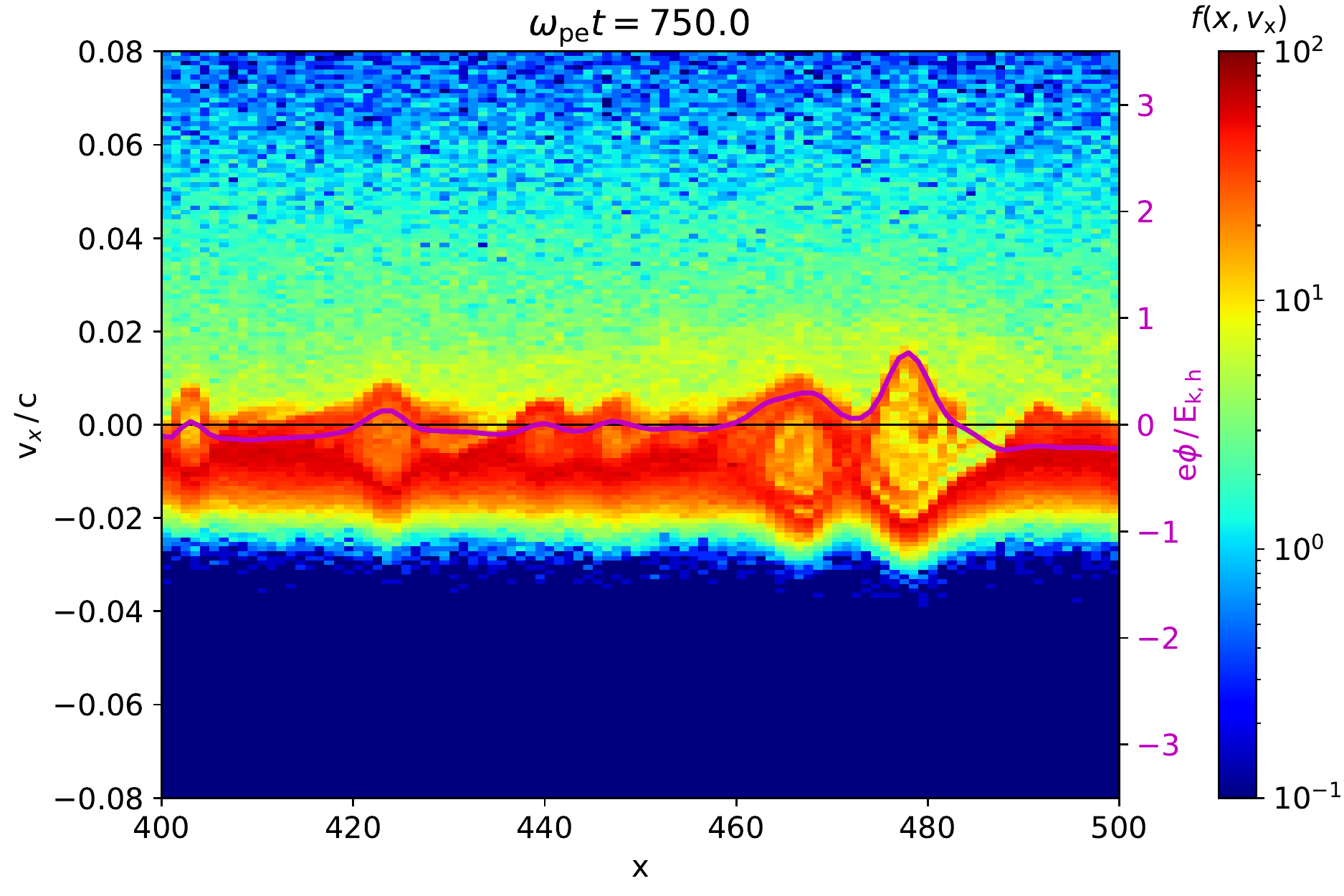}{0.49\textwidth}{(a)}
        \fig{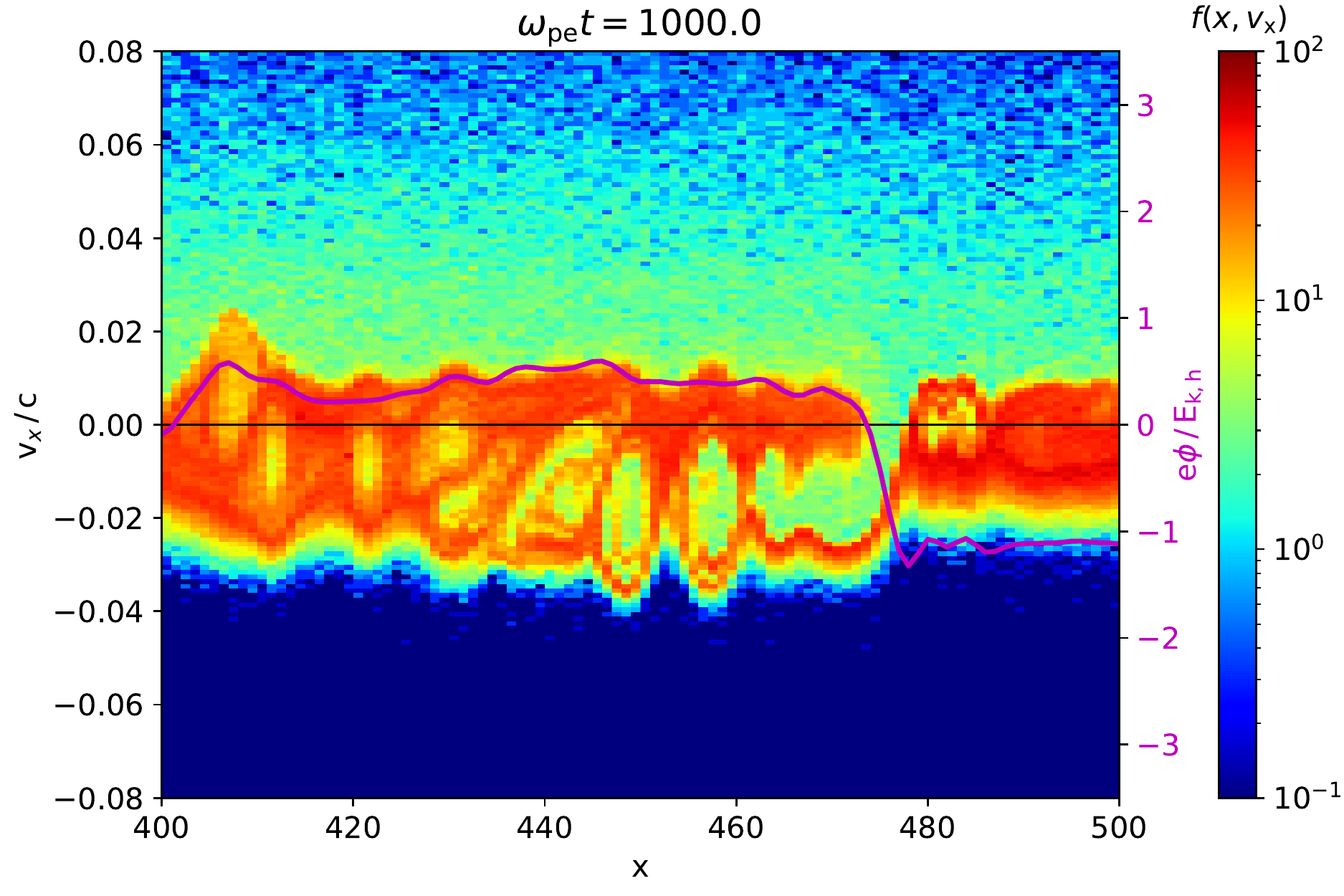}{0.49\textwidth}{(b)}
    }
    \gridline{
        \fig{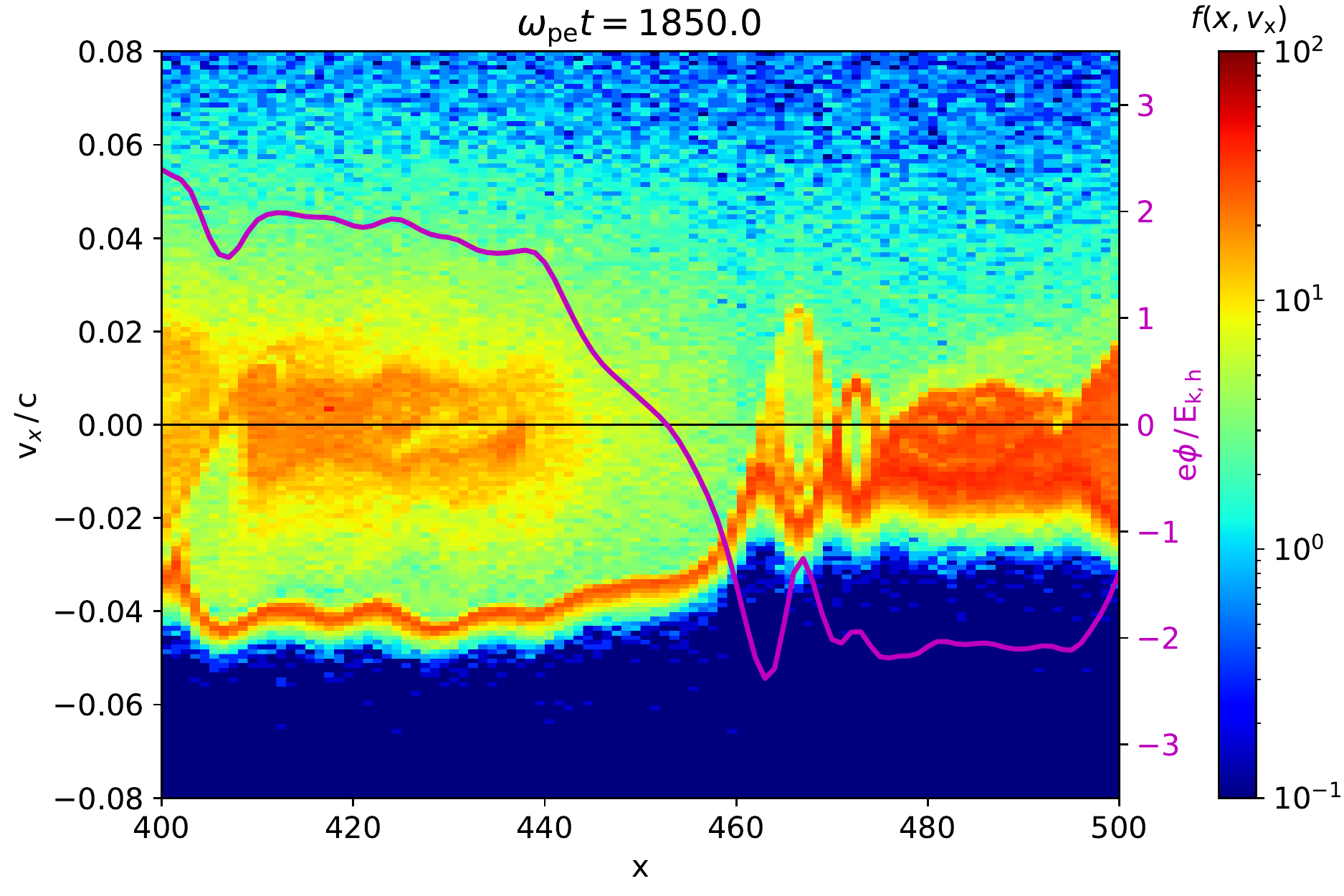}{0.49\textwidth}{(c)}
        \fig{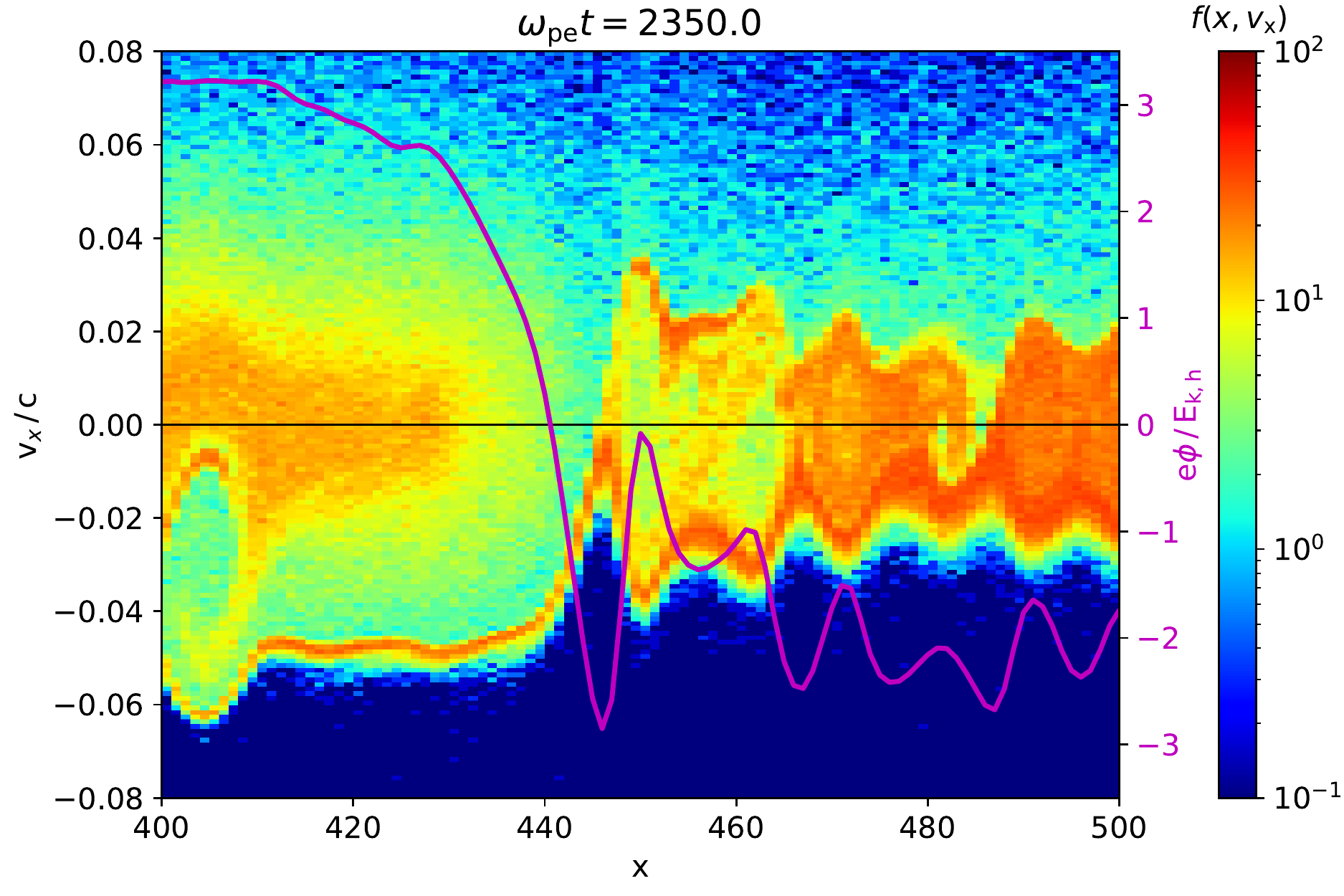}{0.49\textwidth}{(d)}
    }
    \gridline{
        \fig{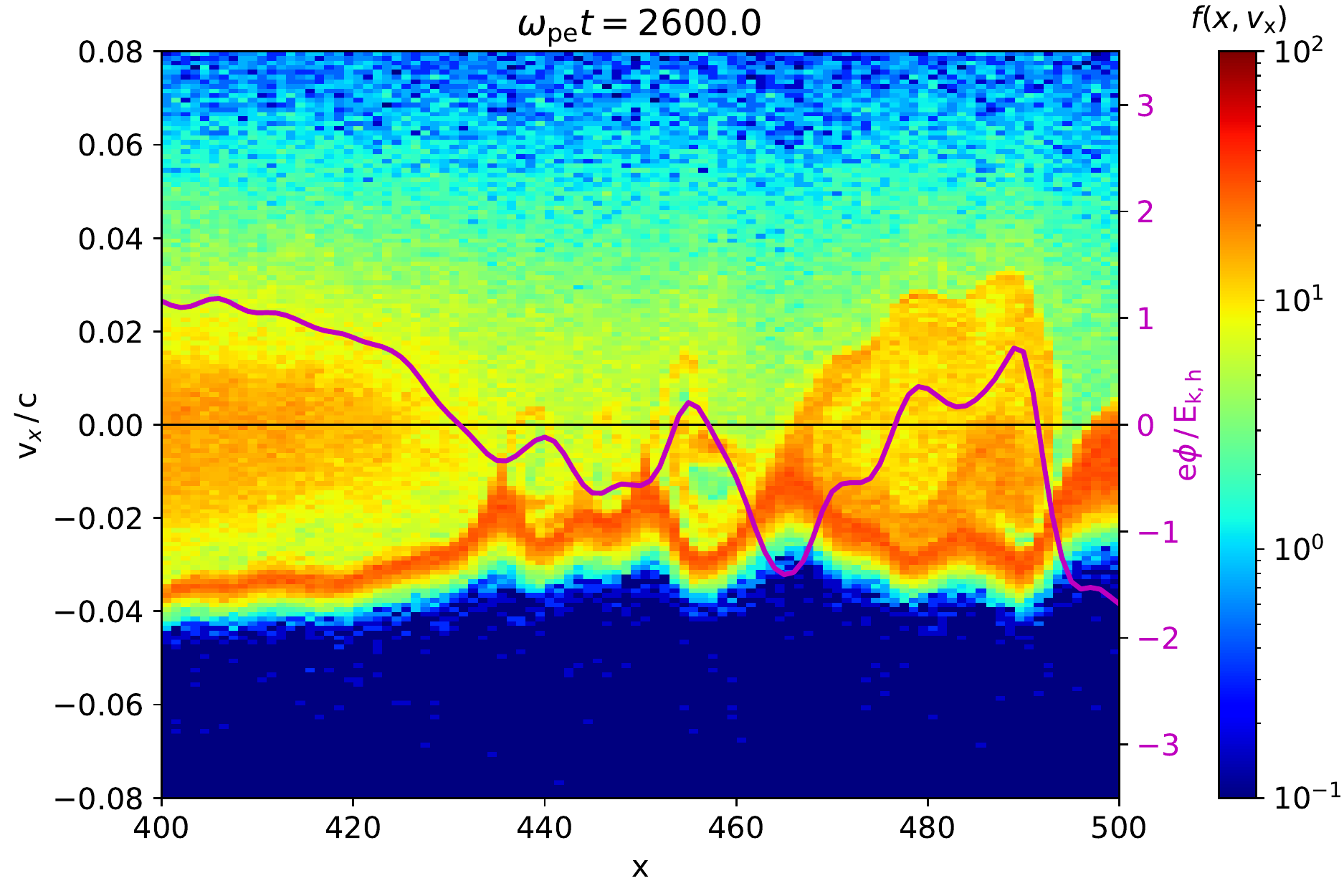}{0.49\textwidth}{(e)}
        \fig{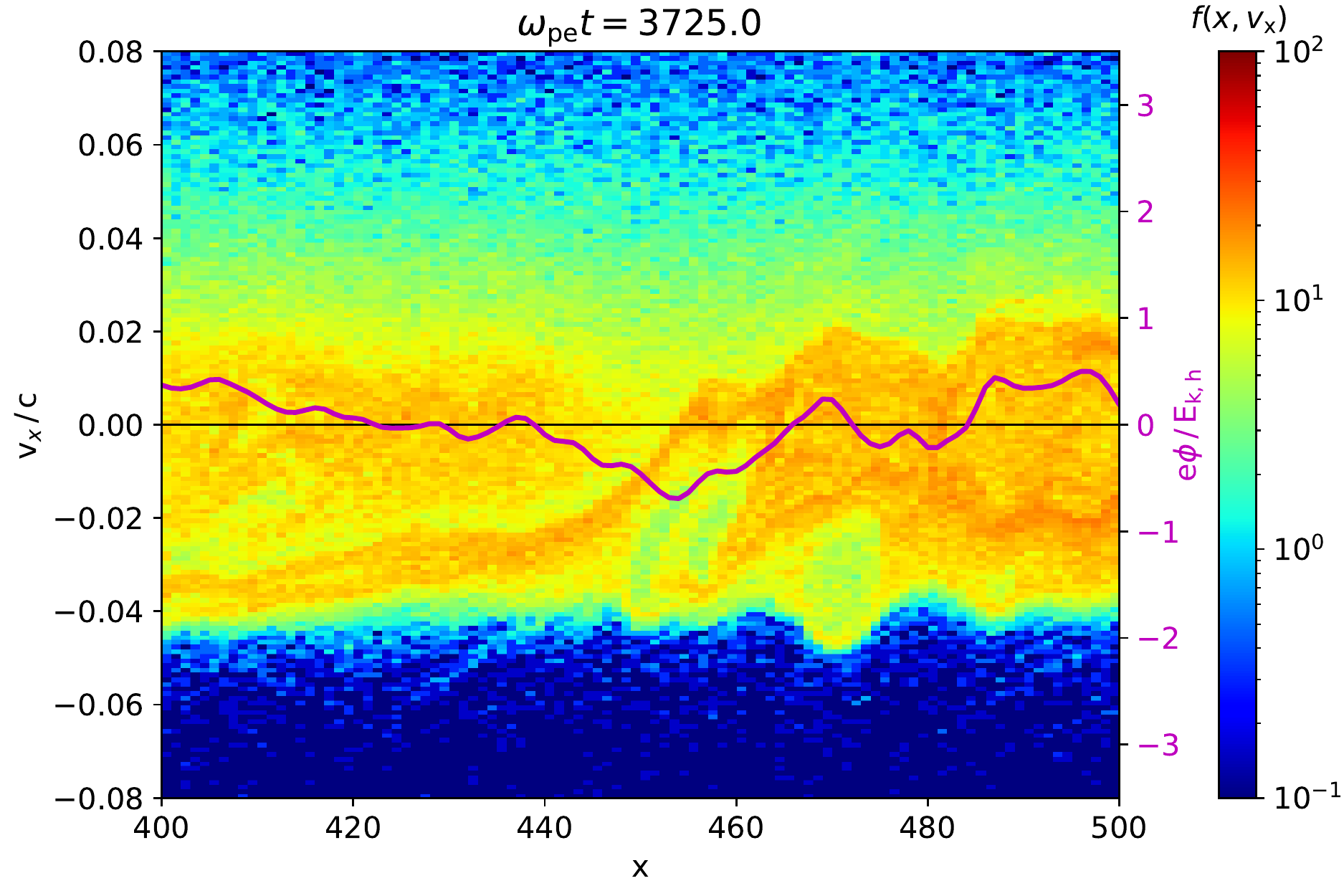}{0.49\textwidth}{(f)}
    }
    \caption{Kappa model: The electron velocity $v_x$ distribution along $x-$ axis for Front 2 in the selected times from the front creation to its dissipation.
    The electric potential at corresponding times is overlaid (magenta line).}
    \label{fig14}
\end{figure*}

Figure~\ref{fig14} shows the evolution of the electron velocity
distribution $v_x$ for Front 2. At time $\omega_\mathrm{pe} t = 750$
 and in the position
$x = 480$, a disturbance is formed. Here the return current (the
maximum of the distribution function is at negative velocities)
increases and the Langmuir turbulence is formed at
1000~$\omega_\mathrm{pe}t$. At this position also the potential jump (DL)
increases until 2350~$\omega_\mathrm{pe} t$ when it is about
$6\,E_\mathrm{k,h}$. The velocity distribution on the left side from DL has two
maxima in this time. The maximum at $v_\mathrm{x} = 0~c$ corresponds to
electrons of the background plasma. The stronger and more narrow maximum
($v_\mathrm{x} =-0.05~c$) is a product of DL that accelerates the electrons
flying from the right of the DL. At this time, on the right side of DL, the
distribution is violated by the beam ($v_\mathrm{x} = 0.025~c$ in the position
$x \sim 450 - 460$) through the beam-plasma instability forming
Langmuir turbulence also on the right side of DL. After this time, the DL
dissipates. The return current decreases and variations of the distribution
together with the electric potential diminish.

\begin{figure*}
    \centering
    \gridline{
        \fig{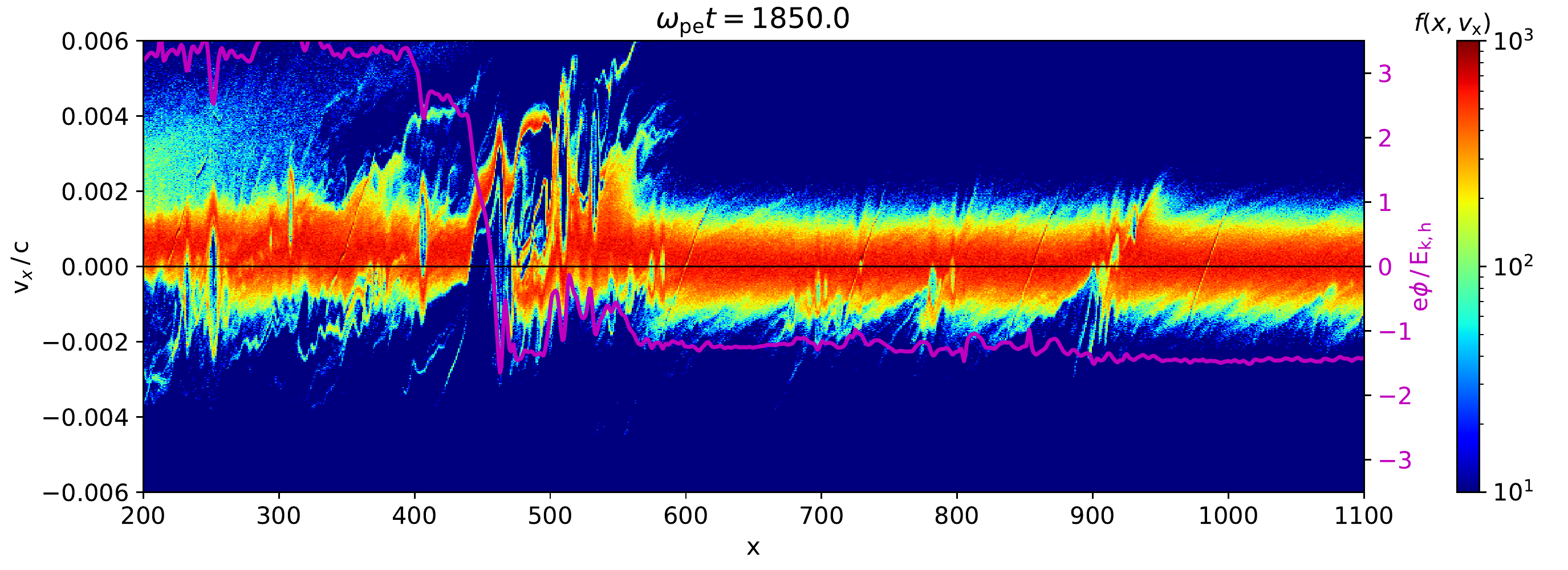}{0.85\textwidth}{(a)}
    }
    \gridline{
        \fig{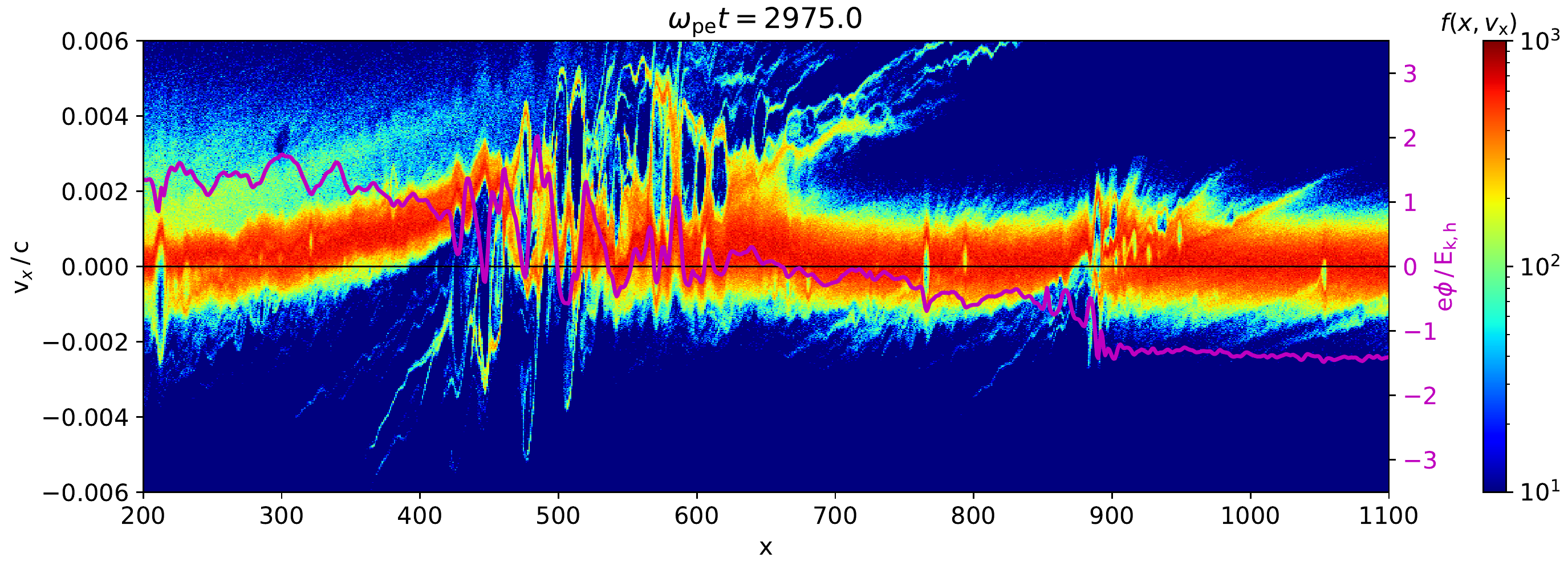}{0.85\textwidth}{(b)}
    }
    \gridline{
        \fig{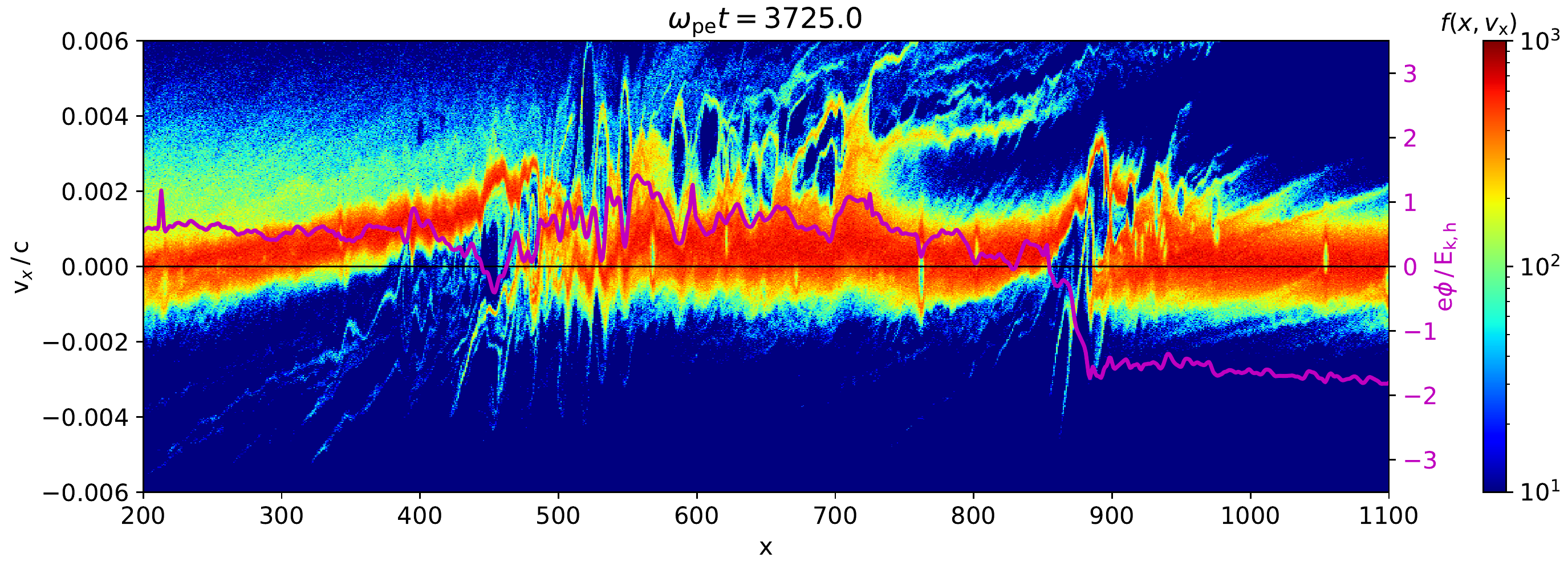}{0.85\textwidth}{(c)}
    }
    \caption{Kappa model: Proton velocity $v_x$ distribution along $x$ axis for Front 2 and 3 and for three selected times $\omega_\mathrm{pe} t =$~1850, 2975, and $3750$. The electric potential is overlaid (magenta line).
    (a) The distribution at time when DL in Front 2 is formed.
    (b) DL of Front 2 is dissipating and generating the proton beam with
    the velocity $v_\mathrm{x} \approx 0.006\,c$ and at locations $x = 720-850$.
    At this time DL ($x = 900$) of Front 3 starts its formation.
    (c) Proton beam from Front 2 disturbs the DL of Front 3. Compare with Figure~\ref{fig4}.}
    \label{fig15}
\end{figure*}

The evolution of the proton velocity $v_x$ distribution is shown in
Figure~\ref{fig15}. First, a disturbance in the velocity distribution can be
seen at $\omega_\mathrm{pe} t = 1850$ and in $x \sim 500$. The protons that
pass the DL from the left to right side are accelerated and gain a high
positive velocity. The protons on the right side of DL and having the negative
velocity are confined on the right side of the DL. Both groups of protons
interact through the beam-beam instability, and their large velocity difference
determines high proton temperatures on the right side of DL (compare the
distribution function with Figure~\ref{fig4} in the same location). The proton
electrostatic turbulence evolves in the position $x = 400-700$
(Figure~\ref{fig15}b) and deforms the DL. A new DL is created at $x = 875$. At
3725~$\omega_\mathrm{pe} t$, the turbulence calms down, potential
smooths, and the released proton beam passes towards Front 3.

\begin{figure*}
    \centering
    \gridline{
        \fig{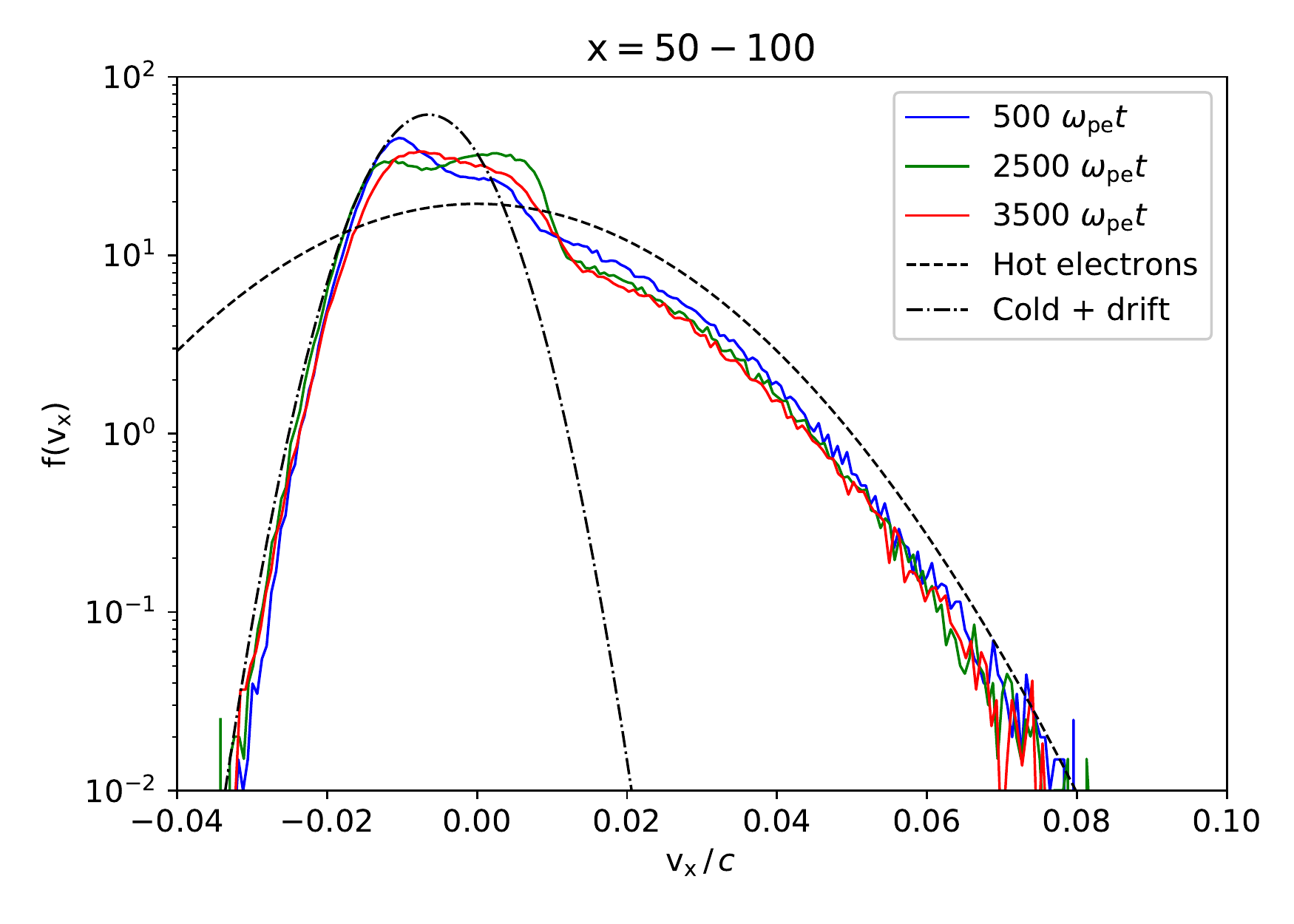}{0.49\textwidth}{(a)}
        \fig{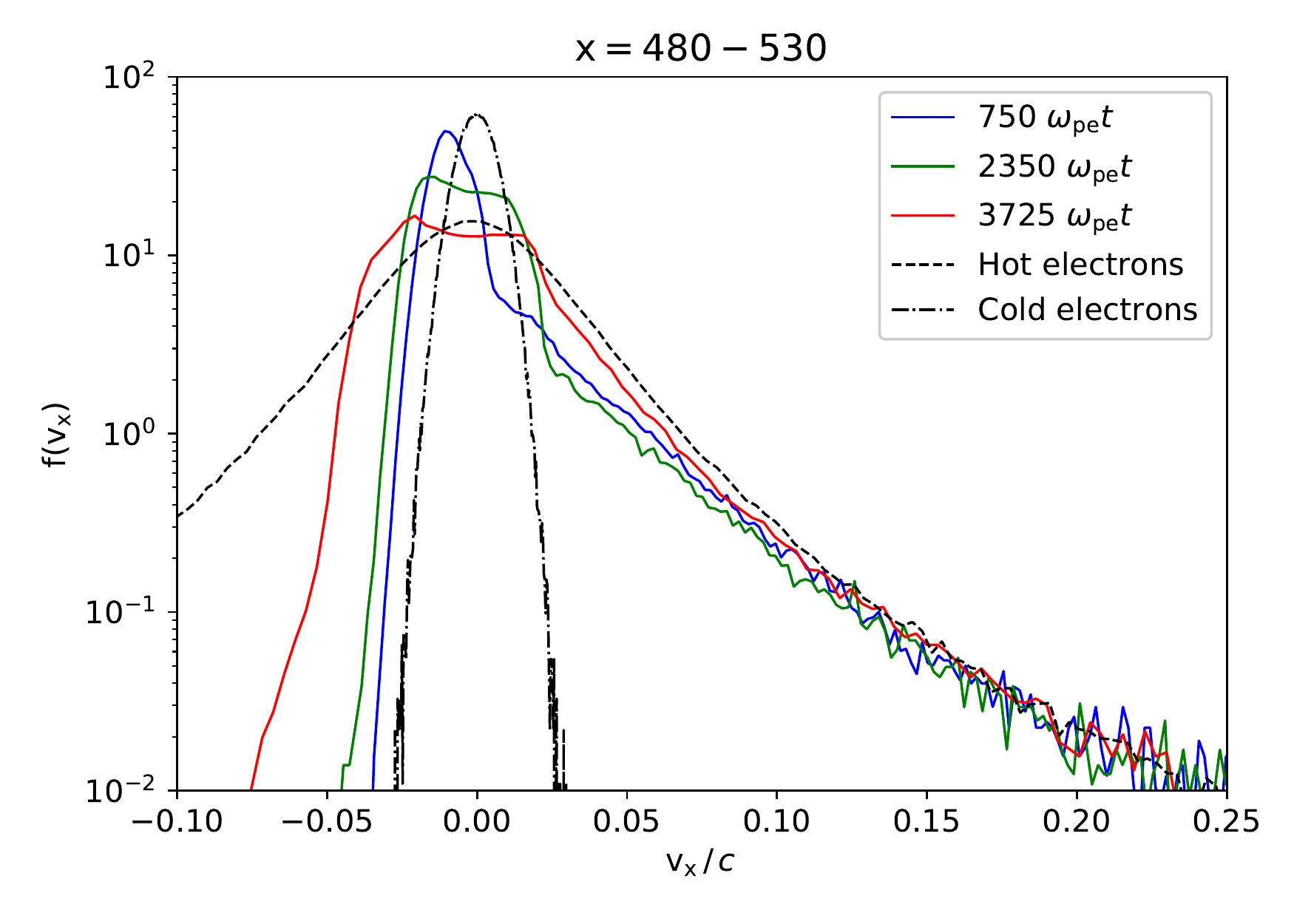}{0.49\textwidth}{(b)}
    }
    \caption{Distribution functions of the electron velocity $v_\mathrm{x}$ on the right side of DLs during their evolution.
    The scales in the horizontal axes in the both figures are different.
    (a) Maxwell model. The distributions in various times calculated in the spatial interval $x=50-100$.
    \textit{Dashed line:} Initial Maxwell distribution of hot electrons for comparison.
    \textit{Dash-dotted line:} Initial Maxwell distribution of cold electrons with the added drift
    velocity caused by the return current.
    (b) Kappa model. The distributions in various times calculated for Front 2 in spatial interval $x=480-530$.
    \textit{Dashed line:} Initian kappa distribution of hot electrons for comparison.}
    \label{fig16}
\end{figure*}

The electron velocity ($v_\mathrm{x}$) distribution functions computed
at different times for the Maxwell and Kappa models are in Figure~\ref{fig16}.
The distributions were computed on the right side of the DL, i.e., in the space
interval of $x=50-100$ for Maxwell model and in space interval of $x=480-530$
for Kappa model, respectively. In both cases, the distributions are asymmetric.
While the parts of distributions with positive velocities are mainly caused by
the electron flux through the DL from the hot part, the parts of the
distributions with negative velocities contain electrons from the cold part of
the simulation. In the Maxwell model (Figure~\ref{fig16}a), after start of the
simulation, the core of the distribution moves to the negative velocities, thus
forming the return current; see the fit of the distribution parts with the
negative velocities by the Maxwell distribution of cold electrons shifted due
to the return current (dash-dotted line). After the transient initial flow of
tail hot electrons, the part of the distribution with positive velocities
remains without significant changes for the whole simulation time (i.e. during
the existence of DL). Also, no significant changes were found in the part of the
distribution with negative velocities. However, the position of the
distribution maximum changes. After the start, when the return current is
formed, the distribution maximum is at negative velocity of about $v = -1.5\,v_\mathrm{he}$.
Then, the part of distribution increases where the velocities are around zero.
At 2500$\,\omega_\mathrm{pe}t$, the maximum of distribution shifts to the
positive velocity of about $v = 0.5\,v_\mathrm{he}$ and then back to negative
velocities at 3500$\,\omega_\mathrm{pe}t$. For comparison in
Figure~\ref{fig16}a we added the initial Maxwell distribution of hot electrons
(dashed line). As can be seen, this distribution in the part of positive
velocities is always greater than the distribution in the right side of the DL.
It corresponds to a reduction of the particle flux and also heat flux in the
$x$ direction.

In Kappa model (Figure~\ref{fig16}b) the evolution of the distribution
on the right side of DL (front 2) is different comparing to  Maxwell model. The
part of the distribution with negative velocities slowly extends to higher
negative velocities, thus increasing the return current. When the DL diminishes
at around 3000$\,\omega_\mathrm{pe}t$, the distribution maximum becomes flatter
and broader. For comparison in Figure~\ref{fig16}a we added the initial kappa
distribution of hot electrons (dashed line). Comparing this distribution with
those on the right side of the DL, it can be seen that at about
750$\,\omega_\mathrm{pe}t$ all distributions for velocities above  $v \sim
0.12�\,c$ are the same. This means that the hot kappa electrons with high
velocities freely propagate through the DL. However, at lower velocities in the
range $v \sim 0.025 - 0.12\,c$ the distributions, expressed by the blue and
green line, i.e., at times of the DL existence, are lower than the the initial
kappa distribution of hot electrons (dashed line). This decrease corresponds to
a reduction of the particle flux and also heat flux in the $x$ direction due to
a presence of the DL. When the DL diminishes at the end of the simulation, the
hot electron distribution (red line) arises to the initial kappa distribution
(dashed line). This means that with the disappearance of DL the particle flux
is without any reduction by the DL.

\section{Discussion and Conclusions}

Motivated by the results of ~\cite{Kasparova09,Oka2013} that the X-ray emission
of flare coronal sources can be explained by the kappa electron distribution,
we studied processes of the plasma confinement in these coronal sources.
The numerical results by \cite{2007PhPl...14j0701R} and analytical
solutions by \cite{2011PhPl...18l2303Y,
2012PhPl...19a2304Y,2012PhPl...19e2301Y} also indicate that the kappa
distribution is natural solution of Langmuir turbulence created in the flare
region. We studied an expansion of the hot plasma to cold one using a
particle-in-cell code in two models: a) Maxwell model with the Maxwellial
distributions in both hot and cold plasmas, and b) Kappa model with the kappa
electron velocity distribution of the hot plasma; others plasma components have
the Maxwellian distributions. Based on observations and in order to emphasize
the kappa case effects, we take $\kappa = 2$.

We compared the results of our Maxwell model with those presented in the papers
by ~\citet{Li2012,Sun19}. The results are very similar despite of some
small differences that are owing to a partly different setup: a) we used the
numerical system divided into two parts with the hot and cold plasma that
differs to the system with the cold-hot-cold parts in the mentioned papers, b)
we considered hot protons in the hot plasma part, and c) we used the isotropic
particle distributions. Therefore, we do not expect a principal
difference in the results of our Kappa model and that with the bi-kappa
distribution anisotropy.

The main result of our study follows from a comparison of the results obtained
in Kappa and Maxwell models. We found that contrary to the Maxwell model, where
one more or less stable thermal front with DL is formed, in the Kappa model, we
recognized a series of thermal fronts associated with DLs. The
differences between the Maxwell and Kappa model are not caused by different
pressures of the hot plasmas, but by the nature of the velocity distribution of
the hot plasma interacting with the DL.

Now, let us summarize our results in more detail:

At the very beginning of the hot-cold plasma interaction, the electrons and
protons from both plasmas are mixing. The hot electrons from the tail of the
distribution of hot plasma that are flying to the right form an electron beam
that generates the return current and electrostatic waves.

The electron beam that is formed from the distribution tail of the kappa
distribution contains more electrons than in the case of the Maxwell model.
Therefore, the electrostatic waves are stronger in the Kappa case.
While in the Maxwell model, these electrostatic waves generate only weak
ion-acoustic waves by the non-linear processes, the stronger electrostatic
(Langmuir) waves in the Kappa model are accumulated at some locations in the
cold plasma region, and by the ponderomotive force, they generate the plasma density
depressions. This process is known as the collapse of Langmuir waves
\citep{Zakharov1972}. These density depressions are then a location for the
formation of the thermal fronts with DL. We found that the maxima of the
electrostatic wave energy are at locations of Front 2 and 3, and the maximum in
Front 2 is higher than that in Front 3.

The front in the Maxwell model and Front 1 in the Kappa model are connected
with a significant jump in the electron temperature. There is a DL at the front
in the Maxwell model, but no significant DL at Front 1 in Kappa model. However,
at Front 2 in the Kappa model, there is a DL that has a higher potential jump
than in the Maxwell model. It is because the tail of the kappa distribution
contains more electrons than Maxwell distribution. The potential jump in the
Maxwell model is about $e\phi = 0.8\,E_\mathrm{k,h}$ (86~eV) which is in agreement with
\cite{Li2012,Li2014}. The potential jump in the Kappa model is higher; its
maximum is about $e\phi = 6\,E_\mathrm{k,h}$ (645~eV).

Front 2 and 3 in the Kappa model can be described more likely as temperature
enhancements, not like the thermal front in the Maxwell model. From both sides,
they are separated by the plasma with a lower plasma temperature. These
temperature enhancements are connected with DLs and density depressions. The
proton temperature enhancement is located on the right side of DL.

The process of the thermal front formation can be described as follows. First,
the hot electron beam from the tail of hot electron distribution generates the
return current and electrostatic waves. Because the return current is unstable by
the Buneman instability, it generates density waves that are progenitors
for the thermal fronts. Then the electron flow creates a potential jump with a
density depression. As the density depression increases, the potential jump
increases, and the double layer with its typical particle flows is generated.

The thermal front formation process occurs independently on the hot-cold
transition width because the electrons from the hot plasma always have higher
velocities than those from the colder plasma and thus overtake them and form
the electron beam. A moment of the beam formation is then the start of the
thermal front formation.

On the other hand, the process of the thermal front dissipation seems to be
connected with the two-stream proton instability (Langmuir turbulence) at the right side of
DL. As the instability reduces the DL jump, the electron flux increases. In the
Maxwell model, the instability is weak and the DL can be reinforced almost in
the instability location. However, in the Kappa model, the instability is
strong enough to suppress the DL reinforcement. The DL is formed in more
distant and colder parts of the model. Both formation and dissipation DL
processes can be repeated until there is a source of the hot plasma.

We can estimate the dissipation time of thermal fronts. Let us suppose
that the width of the density depression is $d = 20\Delta$. The depression is
created by the both electrons and protons. During the front dissipation both
these elements step by step fill the density depression. Because the filling
time by protons is longer than that by electrons and also due to the charge
neutrality, in the estimation, we consider only protons. The proton filling time
is about $d / v_\mathrm{ci} \approx 770\,\omega_\mathrm{pe}t$, where $v_\mathrm{ci} = 6.488\times10^{-4}\,c$ is the cold proton thermal speed. This time agrees with
the dissipation of Front 2 in the time interval
$2500 - 3300\,\omega_\mathrm{pe}t$.

All fronts in the Maxwell and Kappa cases move with the ion-acoustic speed. The
Front 2 moves to the left with velocity $1 \times 10^{-3}\,c$ until
2000~$\omega_\mathrm{pe}t$. Then, during the front dissipation, the motion
direction changes by the electron flow to the right. Front 3 has constant
velocity $-6\times 10^{-4}\,c$.

During the evolution of fronts with the DL, the Langmuir wave packets
and maybe even solitons appear (Figure~\ref{fig12}). It is due to that the
electrostatic energy density of Langmuir waves $W$ exceeded the local thermal
energy density,
\begin{equation}
\frac{W}{nk_\mathrm{B} T} > \frac{k \lambda_\mathrm{D}}{N} \left( \frac{m_\mathrm{e}}{m_\mathrm{i}} \right)^{\frac{1}{2}}
\label{eq3}
\end{equation}
where $N$ is the number of particles in Debye's sphere
\citep{1975ZhETF..69..155Z,1984RvMP...56..709G}, $k$ is the wave vector. If we
assume that the typical soliton wave vector $k$ is independent on the ion mass,
the increase of ratio $m_\mathrm{i}/m_\mathrm{e}$ to natural values results in
decreasing of the strong Langmuir turbulence threshold on right side of
Equation~\ref{eq3}. Therefore, for natural values of
$m_\mathrm{i}/m_\mathrm{e}$ it is expected to have more generated solitons than
in our case. Also, the dissipation should be more difficult.

We analyzed the electron velocity distributions just on the right side
of DL for the Maxwell model and Kappa model. In both cases the distributions
are asymmetric owing to the expansion of hot plasma electrons into the region
with cold electrons. We found that during an existence of the DL the
distribution on the right side of the DL, formed mainly by the hot electrons
propagating through DL, is lower than the initial distribution of hot electrons
in the hot plasma region. It indicates a reduction of the hot particle (or
heat) flux through DL.

Our simulations show that in the expansion of the hot plasma into a cold one,
in the both Kappa and Maxwell models, electrostatic and ion-acoustic waves are
generated. Considering a coalescence of these waves into the electromagnetic
waves, described processes can be detected in the solar radio emission. We
propose that the electrons from the hot-plasma distribution tail can generate
some type III-like bursts. On the other hand, owing to a motion of the thermal
front, some slowly drifting bursts can be observed in the dynamic radio
spectra.

\acknowledgements We acknowledge support from Grants 18-09072S, 19-09489S,
20-09922J, and 20-07908S of the Grant Agency of the Czech Republic. This work
was supported by The Ministry of Education, Youth and Sports from the Large
Infrastructures for Research, Experimental Development and Innovations project
``IT4Innovations National Supercomputing Center -- LM2015070''. Computational
resources were provided by the CESNET LM2015042 and the CERIT Scientific Cloud
LM2015085, provided under the programme ``Projects of Large Research,
Development, and Innovations Infrastructures''.

\end{document}